\documentclass[english,nofootinbib,showpacs,preprint,preprintnumbers,aps]{revtex4}
\usepackage[latin9]{inputenc}
\usepackage{color}
\usepackage{babel}
\usepackage{mathtools}
\usepackage{amsmath}
\usepackage{amssymb}
\usepackage{wasysym}
\usepackage{mathdots}
\usepackage{graphicx}
\usepackage[unicode=true,
 bookmarks=true,bookmarksnumbered=false,bookmarksopen=false,
 breaklinks=false,pdfborder={0 0 1},backref=page,colorlinks=true,pdfpagemode=FullScreen]
 {hyperref}
\hypersetup{pdftitle={Fermionic DM and neutrino masses in BL}}

\makeatletter
\@ifundefined{textcolor}{}
{%
 \definecolor{BLACK}{gray}{0}
 \definecolor{WHITE}{gray}{1}
 \definecolor{RED}{rgb}{1,0,0}
 \definecolor{GREEN}{rgb}{0,1,0}
 \definecolor{BLUE}{rgb}{0,0,1}
 \definecolor{CYAN}{cmyk}{1,0,0,0}
 \definecolor{MAGENTA}{cmyk}{0,1,0,0}
 \definecolor{YELLOW}{cmyk}{0,0,1,0}
}

\usepackage{babel}

\usepackage{amsfonts}
\usepackage{epsfig}
\usepackage{bm}

\usepackage{hyperref}
\hypersetup{
     colorlinks = true,
     linkcolor = blue,
     urlcolor  = violet,
     citecolor = magenta,
     anchorcolor = cyan,
}

\usepackage{graphicx}

\usepackage{verbatim}
\usepackage{ae,aecompl}
\usepackage{amsmath,amssymb,mathrsfs}

\newcommand{\head}[1]{\textnormal{\textbf{#1}}}

\definecolor{darkpastelpurple}{rgb}{0.59, 0.44, 0.84}
\definecolor{frenchlilac}{rgb}{0.53, 0.38, 0.56}
\definecolor{violet}{rgb}{0.56, 0.0, 1.0}

\providecommand{\ZZ}{\mathbb{Z}}
\providecommand{\lag}{\mathscr{L}}

\usepackage{capt-of}

\makeatletter
\renewcommand*{\@fnsymbol}[1]{\ifcase#1 \or \knight  \or \queen \else \rook \fi}
\makeatother

\usepackage{chessfss}
\usepackage{adjustbox}

\setfigstyle{\itshape}

\providecommand{\knight}[1][2.5ex]{
\adjustbox{Trim=2.3pt 2.35pt 2.5pt 0pt,width=#1,raise=-0.3ex,margin=0.14ex 0ex 0.14ex
0ex}{\BlackKnightOnWhite}
}

\makeatother

\begin{document}

\title{Fermionic dark matter and neutrino masses in a $\cal{B}-\cal{L}$
model}

\author{B. L. Sánchez--Vega}

\email{brucesanchez@anl.gov (brucesanchez@gmail.com)}

\affiliation{Argonne National Laboratory, 9700 S. Cass Avenue, Argonne, IL 60439. }

\author{E. R. Schmitz}

\email{ernany@th.physik.uni-bonn.de}

\affiliation{Bethe Center for Theoretical Physics and Physikalisches Institut,
Universität Bonn, Nussallee 12, D-53115 Bonn, Germany.}
\begin{abstract}
In this work we present a common framework for neutrino mass and dark
matter. Specifically, we work with a local $\mathcal{B-L}$ extension
of the standard model which has three right-handed neutrinos, $n_{R_{i}}$,
and some extra scalars, $\Phi,\,\phi_{i}$ besides the standard model
fields. The $n_{R_{i}}$'s have non-standard $\mathcal{B-L}$ quantum
numbers and thus these couple to different scalars. This model has
the attractive property that an almost automatic $\ZZ_2$ symmetry
acting only on a fermionic field, $n_{R3}$, is present. Taking advantage
of this $\ZZ_2$ symmetry, we study both the neutrino mass generation
via a natural see-saw mechanism in low energy and the possibility
of $n_{R3}$ to be a DM candidate. For this last purpose, we study
its relic abundance and its compatibility with the current direct
detection experiments. 

\end{abstract}
\maketitle

\section{introduction}

\label{sec:introduction}At least two experimental evidences demand
for physics beyond the standard model (SM). The first one comes from
the well stablished neutrino oscillation experiments \citep{SuperK,SNO,KamLAND,MINOS}
which imply that all the three known neutrinos ($\nu_{e},\,\nu_{\mu},\,\nu_{\tau}$)
are quantum superpositions of three massive states $\nu_{i}$ ($i=1,\,2,\,3$).
The second evidence is firmly stablished from several observations
and studies of gravitational effects on different scales, which points
out that most of the Universe's mass consists of non-baryonic dark
matter (DM). Specifically, the Planck collaboration has determined
that the DM relic abundance is given $\Omega_{\text{DM}}h^{2}=0.1193\pm0.0014$
\citep{PLANCK_2013}. 

In order to explain both of these evidences, it is now clear that
the SM has to be extended. In the neutrino case, usually new fermionic
fields, $n_{R}$'s, are introduced to generate Dirac mass terms for
neutrinos. The $n_{R}$ fields are, in general, singlet under the
SM gauge groups, and thus these can also have Majorana mass terms.
Moreover, in order to explain the smallness of active neutrino masses,
the $n_{R}$'s usually get large masses via the well known see-saw
mechanism \citep{KingLuhn:2013neutrinomixing,see-saw_Gonzalez_Garcia}.
On the other hand, the existence of DM in the Universe requires at
least a new massive particle since SM does not provide any viable
DM candidate. The most studied and well motivated candidates for DM
are the weakly interacting massive particles (WIMPs). In general,
WIMPs are neutral, stable and are present in a plethora of extensions
of the SM \citep{WIMPS1,WIMPS2,WIMPS3,WIMPS4}. Nowadays, there are
several astroparticle experiments actively pursuing detection of WIMP
DM candidates in direct and indirect ways. The direct detection experiments
\citep{LUX_2014,XENON100_2012,SuperCDMS_2014} have set upper bounds
on WIMP-nucleon elastic scattering, whereas the indirect ones \citep{ID1,ID2,ID3,ID4,ID5}
presented upper limits on the thermal average of the same scattering
cross section $\left\langle \sigma v_{\textrm{Mol}}\right\rangle $.

In this work we study a scenario that simultaneously offers an explanation
for the previously mentioned open questions of the SM. In special,
we present a local $\mathcal{B-L}$ extension of the SM in which there
are three $n_{R}$ fermionic fields and some extra scalars, $\Phi,\,\phi_{i}$.
The $\mathcal{B-L}$ quantum numbers of the extra new fermionic fields
come from exotic solutions of anomaly constraints. These solutions
were found for the first time in Ref. \citep{Montero:2007cd}. Here,
we propose a simplified version of the model in Ref. \citep{Montero:2007cd,Bruce:2011Neutrinos}
where an almost natural $\ZZ_2$ symmetry stabilizes $n_{R_{3}}$
and thus allows it to be a DM candidate. Another appealing feature
of this model is that it implements a see-saw mechanism at low energy
because neutrino masses are proportional to $V_{\Phi}^{2}/V_{\phi}$
(where $V_{\Phi}$ and $V_{\phi}$ are VEVs of the $\Phi,$ $\phi_{i}$,
respectively) and $V_{\Phi}$ has been set in MeV energy scale. The
discrete symmetry also simplifies the task of setting the Yukawa couplings
in the neutrino mass Lagrangian in order to agree with the neutrino
oscillation parameters. There have been extensive studies on these
two matters, in special in gauge extensions of the SM (a few of which
are contained in the Refs. \citep{Bruce:2014DM,DM_with_neutrinos1,DM_with_neutrinos2}).
This is so because U$(1)$ gauge factors are contained in grand unification
theories \citep{GUT1,GUT2}, supersymmetric models \citep{SUSY_add_1}
and left-right models \citep{left_right_1,left_right_2}. 

The paper is organized as follows. We start by discussing the model
in Sec. \ref{sec:model}. In that section we present its field content
and general Lagrangian. We also show the almost natural $\ZZ_{2}$
symmetry in the model which stabilizes the DM candidate $n_{3R}$.
In Sec. \ref{sec:Scalar Sector} we study the scalar sector in detail.
We obtain analytical formulas for both the mass eigenstates and the
eigenvalues when it is possible. We also include a discussion about
the consequences of the presence of the Majoron $J$ which is due
to the breaking of an accidental global U$(1)_{J}$ symmetry. Specifically,
we show that it escapes the current bounds on energy loss in stars
\citep{geJ_stars1,geJ_stars2}, effective number of neutrinos $N_{\textrm{eff}}$
\citep{PLANCK_2013}, and the invisible decay widths of Higgs \citep{Ellis_2013,Giardino2014,Atlascollaboration,CMScollaboration}
and $Z_{1}$ gauge boson. In Sec. \ref{sec:Neutrino Masses} we analytically
find the parameters of the neutrino mass matrices in order to satisfy
the data from the neutrino oscillation \citep{Olive:2014PDG} and
other constraints such as lepton flavor violation (LFV) \citep{mee_future,mee_CUORE_2015};
the sum of the SM neutrinos masses \citep{PLANCK_2013}; and effective
Majorana mass $m_{ee}$ from double beta decay experiments \citep{mee_future,mee_CUORE_2015}.
In Sec. \ref{sec:dark_matter} we carry out a study of the relic dark
matter abundance and the direct detection prospects. A general discussion
follows in Sec. \ref{sec:Conclusions} where we present our conclusions.
Finally, in Appendix \ref{Appendix} we show the general minimization
conditions coming from the scalar potential.

\section{the model}

\label{sec:model} We consider an extension of the SM based on the
gauge symmetry SU$(2)_L\otimes$U$(1)_{Y^{'}}\otimes$U$(1)_{\cal{B}-\cal{L}}$
where $\cal{B}$ and $\cal{L}$ are the usual baryonic and leptonic
numbers, respectively, and $Y^{\prime}$ is a new charge different
from the hypercharge $Y$ of the SM. The values of $Y^{\prime}$ are
chosen to obtain the hypercharge $Y$ through the relation $Y=[Y^{\prime}+{\cal {B}-{\cal {L}}}]$,
after the first spontaneous symmetry breaking. The fields of this
model with their respective charges are shown in Table \ref{table1}.
Actually, this model is a simplified variation of the one introduced
in Refs. \citep{Montero:2007cd,Bruce:2011Neutrinos}. Specifically,
here we have removed one of the extra doublets of scalars considered
there. As we will show below, this allows an almost automatic $\ZZ_2$
symmetry that stabilizes the DM candidate, $n_{R3}$. The remaining
scalar fields are enough to give mass to the neutrinos at tree level.
It is also important to note that there is an exotic charge assignment
for the $\ensuremath{{\cal {B}-{\cal {L}}}}$ charges where $(\ensuremath{{\cal {B}-{\cal {L}}}}){}_{n_{R1},n_{R2}}=-4$
and $(\ensuremath{{\cal {B}-{\cal {L}}}}){}_{n_{R3}}=5$ different
from the usual one where $(\ensuremath{{\cal {B}-{\cal {L}}}}){}_{n_{Ri}}=1$
with $i=1,\,2,\,3$.\begin{table}[h]
\begin{tabular}{||c||ccc||c||ccc||}  
\hline
\hline  
\head{Fermion} & $I_3$ &  $Y'$ & $\cal{B}-\cal{L}$ & \head{Scalar} & $I_3$ &  $Y'$ & $\cal{B}-\cal{L}$ \\ 
\hline 
$\nu _{eL},\,e_{L}$ & $\pm$ 1/2 & 0  & -1 &$H^{+,0}$ & $\pm$ 1/2 & 1 & 0 \\ 
$e_{R}$ &  0& -1& -1 & $\Phi^{0,-}$ & $\pm$ 1/2 & -4 & 3   \\ 
$u_{L},\,d_{L}$ & $\pm$ 1/2 & 0   & 1/3 & $\phi_{1}$ & 0 & -8 & 8    \\
$u_{R}$  & 0 & 1   & 1/3 & $\phi_{2}$ & 0 & 10 & -10   \\
$d_{R}$ & 0  & -1  & 1/3 & $\phi_{3}$ & 0 & 1 & -1 \\
$n_{R1}, n_{R2}$ & 0  & 4 & -4 & $\phi_{\textrm{X}}$ & 0 & 3 & -3  \\
$n_{R3}$  & 0  & -5 & 5 &  & &  &  \\
\hline
\hline 
\end{tabular}
\caption{Quantum number assignment for the fields in the model. $I_3$, $Y'$  and $\cal{B}-\cal{L}$ are the quantum numbers under the symmetry groups SU$(2)_L$, U$(1)_{Y^{'}}$, U$(1)_{\cal{B}-\cal{L}}$, respectively.} 
\label{table1} 
\end{table}

With the field content in Table \ref{table1}, we can write respectively
the most general renormalizable Yukawa Lagrangian and scalar potential
respecting the gauge invariance as follows
\begin{eqnarray}
-\lag_{\text{Y}} & = & Y_{i}^{(l)}\overline{L}_{Li}e_{Ri}H+Y_{ij}^{(d)}\overline{Q}_{Li}d_{Rj}H+Y_{ij}^{(u)}\overline{Q}_{Li}u_{Rj}\widetilde{H}+\mathcal{D}_{im}\overline{L}_{Li}n_{Rm}\Phi\nonumber \\
 &  & +\frac{1}{2}\mathcal{M}_{mn}\overline{(n_{Rm})^{c}}n_{Rn}\phi_{1}+\frac{1}{2}\mathcal{M}_{33}\overline{(n_{R3})^{c}}n_{R3}\phi_{2}+\frac{1}{2}\mathcal{M}_{m3}\overline{(n_{Rm})^{c}}n_{R3}\phi_{3}+\textrm{H.c.},\label{lyukawa}
\end{eqnarray}
and
\begin{eqnarray}
V_{\ensuremath{{\cal {B}-{\cal {L}}}}} & = & -\mu_{H}^{2}H^{\dagger}H+\lambda_{H}\left|H^{\dagger}H\right|{}^{2}-\mu_{\Phi}^{2}\Phi^{\dagger}\Phi+\lambda_{\Phi}\left\vert \Phi^{\dagger}\Phi\right\vert ^{2}-\mu_{\alpha}^{2}\left\vert \phi_{\alpha}\right\vert ^{2}+\lambda_{\alpha}\left\vert \phi_{\alpha}^{\ast}\phi_{\alpha}\right\vert ^{2}\nonumber \\
 &  & +\kappa_{H\Phi}\left\vert H\right\vert ^{2}\left\vert \Phi\right\vert ^{2}+\kappa_{H\Phi}^{\prime}(H^{\dagger}\Phi)(\Phi^{\dagger}H)+\kappa_{H\alpha}\left\vert H\right\vert ^{2}\left\vert \phi_{\alpha}\right\vert ^{2}+\kappa_{\Phi\alpha}\left\vert \Phi\right\vert ^{2}\left\vert \phi_{\alpha}\right\vert ^{2}\nonumber \\
 &  & +\kappa_{\alpha\beta}(\phi_{\alpha}^{\ast}\phi_{\alpha})(\phi_{\beta}^{\ast}\phi_{\beta})+\left[\kappa_{123}\phi_{1}\phi_{2}(\phi_{3}^{\ast})^{2}-i\kappa_{H\Phi X}\Phi^{T}\tau_{2}H\phi_{X}+\kappa_{123X}(\phi_{X}^{\ast}\phi_{1})(\phi_{2}\phi_{3})\right.\nonumber \\
 &  & \left.+\kappa'_{3X}(\phi_{X}^{\ast}\phi_{3}^{3})+\textrm{H.c.}\right],\label{potential}
\end{eqnarray}
where $i,j=1,2,3$ are lepton/quark family numbers; $m,n=1,2$; $\widetilde{H}=i\tau_{2}H^{\ast}$
($\tau_{2}$ is the Pauli matrix), and $\alpha,\beta=1,2,3,X$ with
$\alpha\neq\beta$ in the $\kappa_{\alpha\beta}(\phi_{\alpha}^{\ast}\phi_{\alpha})(\phi_{\beta}^{\ast}\phi_{\beta})$
terms. Also, we have omitted summation symbols over repeated indices.

Before we go further, two important remarks are in order. Firstly,
from Eqs. (\ref{lyukawa}) and (\ref{potential}) we see that apart
from the $\frac{1}{2}\mathcal{M}_{m3}\overline{(n_{Rm})^{c}}n_{R3}\phi_{3}+\textrm{H.c.}$
terms, the Lagrangian is invariant under a $\ZZ_{2}$ symmetry acting
in a non-trivial way on the $n_{R3}$ field, i.e. $\ZZ_{2}(n_{R3})=-n_{R3}$
(the rest of fields being invariant under this symmetry). We will
consider the case of this $\ZZ_{2}$ symmetry throughout this work.
Hence, the $n_{R3}$ fermionic field will be the DM candidate. Secondly,
from Eq. (\ref{lyukawa}) we see that quarks and charged leptons obtain
masses just from the $H$ vacuum expectation value, $\left\langle H^{0}\right\rangle \equiv V_{H}$.
Therefore, the $H$ interactions with quarks and charged leptons are
diagonalized by the same matrices as the corresponding mass matrices.
In this case the neutral interactions are diagonal in flavor and there
is no flavor-changing neutral current in the quark and charged lepton
sector. This feature remains after the symmetry basis is changed to
mass basis \citep{Glashow1958,Cheng1987}. However, lepton flavor
violation (LFV) processes coming from the terms proportional to $\mathcal{D}_{im}$
and $\mathcal{D}_{i3}$ can occur at one loop. We will discuss these
processes in more detail in Sec. \ref{sec:Neutrino Masses}.

\section{scalar sector}

\label{sec:Scalar Sector}In the general case this model has a rich
scalar spectrum and its vacuum structure can take several configurations.
However, we are going to make some simplifying and reasonable assumptions
that allow us, in most cases, to obtain analytical formulas in both
the neutrino and the dark matter sectors. We will discuss systematically
our assumptions throughout this paper.

Firstly, as result of the absence of one of the extra doublets and
of writing only the renormalizable terms in the scalar potential,
the model here considered has a Majoron, $J$, in its scalar spectrum.
This is a general conclusion and does not depend on any particular
choice of the set of parameters. Once the neutral scalars develop
non-vanishing vacuum expectation values, VEVs, and using the usual
shifting $\varphi^{0}=\frac{1}{\sqrt{2}}\left(V_{\varphi}+\textrm{Re}\,\varphi+i\,\textrm{Im}\,\varphi\right)$
for the scalar fields (the superscript ``$0$'' means we are taking
the neutral part of the field), we find that $J$ can be written as
\begin{eqnarray}
J & = & \frac{1}{N_{J}}\left[-9\sqrt{2}V_{H}V_{\phi}^{3}\epsilon^{2}\,\textrm{Im}\, H^{0}-9\sqrt{2}V_{H}^{2}V_{\phi}^{2}\epsilon\,\textrm{Im}\,\Phi^{0}\right.\nonumber \\
 &  & \left.+\frac{1}{\sqrt{2}}V_{\phi}^{2}\left(10V_{H}^{2}+\left(3V_{H}^{2}+10V_{\phi}^{2}\right)\epsilon^{2}\right)\,\textrm{Im}\,\phi_{1}+\frac{V_{\phi}^{2}\left(2V_{H}^{2}-\left(3V_{H}^{2}-2V_{\phi}^{2}\right)\epsilon^{2}\right)}{\sqrt{2}}\,\textrm{Im}\,\phi_{2}\right.\nonumber \\
 &  & \left.+3\sqrt{2}V_{\phi}^{2}\left(V_{H}^{2}+V_{\phi}^{2}\epsilon^{2}\right)\,\textrm{Im}\,\phi_{3}+9\sqrt{2}V_{\phi}^{2}\left(V_{H}^{2}+V_{\phi}^{2}\epsilon^{2}\right)\,\textrm{Im}\,\phi_{X}\right],\label{majoron}
\end{eqnarray}
where $N_{J}\equiv V_{\phi}^{2}\sqrt{\left(4V_{H}^{2}+\left(3V_{H}^{2}+4V_{\phi}^{2}\right)\epsilon^{2}\right)\left(58V_{H}^{2}+\left(3V_{H}^{2}+58V_{\phi}^{2}\right)\epsilon^{2}\right)}$
and $\epsilon\equiv V_{\Phi}/V_{\phi}$. We also have defined the
VEVs as $\left\langle \phi_{i}\right\rangle \equiv V_{\phi_{i}}$
with $i=H,\,\Phi,\,\phi_{1},\,\phi_{2},\,\phi_{3},\,\phi_{X}$ and
set $V_{\phi_{1}}=V_{\phi_{2}}=V_{\phi_{3}}=V_{\phi_{X}}\equiv V_{\phi}$
for simplicity. The parameter $\epsilon$ is chosen $\ll1$ as we
will show below.

The presence of $J$ in the physical spectrum is due to an extra symmetry
in the scalar potential in Eq. (\ref{potential}). In other words,
the scalar potential actually has a larger global SU$(2)_L\otimes$U$(1)_{Y^{'}}\otimes$U$(1)_{\cal{B}-\cal{L}}\otimes$U$(1)_J$
symmetry. The last symmetry, U$\left(1\right)_{J}$, acts on the scalar
fields $H,\,\Phi,\,\phi_{1},\,\phi_{2},\:\phi_{3},\,\phi_{X}$ with
charges $-\frac{18}{23},\,-\frac{18}{23},\,1,\,\frac{1}{23},\,\frac{12}{23},\,\frac{36}{23}$,
respectively. We have normalized the charges in order to set the $\phi_{1}$
charge equal to $1$. Also, note that the global symmetry U$\left(1\right)_{J}$
is independent on the U$(1)_{Y^{'}}$ and U$(1)_{\cal{B}-\cal{L}}$
symmetries. This is necessary to consider it as an actual extra symmetry.
Furthermore, U$\left(1\right)_{J}$ can be extended to the total Lagrangian
acting on the fermions $Q_{L},\,u_{R},\,d_{R},\,L_{L},\,e_{R},\,n_{Rm},\,n_{3R}$ 
with charges $0,-\frac{18}{23},\,\frac{18}{23},\,-\frac{59}{23},\,-\frac{1}{2},\,-\frac{1}{2},\,-\frac{1}{46}$,
respectively. Therefore, $J$ is a true Majoron with mass equal to
zero at all orders in perturbation theory. Gravitational effects can
break this symmetry, and thus give mass to the Majoron \citep{Gilbert1989159,Holman1992132,Kamionkowski1992137,Akhmedov_1993,Bruce:2011PQ}.
Studies taking into account these effects on $\mathcal{B-L}$ symmetry
constrain the energy scale of its breakdown to be $<10\,\textrm{TeV}$
\citep{Akhmedov_1993}. However, we are not going to consider this
case. 

The major challenge to models with a Majoron comes from the energy
loss in stars through the process $\gamma+e^{-}\rightarrow e^{-}+J$.
This process is used to put limits on the $\bar{e}eJ$ coupling, $g_{\bar{e}eJ}$,
and it is found that it must be $g_{\bar{e}eJ}\leq10^{-10}$ for the
Sun, and $g_{\bar{e}eJ}\leq10^{-12}$ for the red-giant stars \citep{geJ_stars1,geJ_stars2}.
In our case, $g_{\bar{e}eJ}=\frac{Y_{e}^{(l)}}{\sqrt{2}}\frac{9\sqrt{2}V_{H}V_{\phi}^{3}}{N_{J}}\epsilon^{2}=\frac{m_{e}}{V_{H}}\frac{9\sqrt{2}V_{H}V_{\phi}^{3}}{N_{J}}\epsilon^{2}$
where $Y_{e}^{(l)}$ and $m_{e}$ are the electron Yukawa coupling
to the $H$ scalar and electron mass, respectively. Since $\epsilon=V_{\Phi}/V_{\phi}$,
$V_{\textrm{SM}}=\sqrt{V_{H}^{2}+V_{\Phi}^{2}}$ and $V_{H}\simeq V_{\textrm{SM}}$
(the $\textrm{Re}\, H^{0}$ is the only field giving mass to the top
quark at tree level), we have that $\epsilon\ll1$. Thus, expanding
$g_{\bar{e}eJ}$ in series of $\epsilon$, it is straightforward to
see that $g_{\bar{e}eJ}\simeq\frac{9m_{e}V_{\phi}}{2\sqrt{29}V_{H}^{2}}\epsilon^{2}+{\cal O}(\epsilon^{4})$.
Choosing $V_{\phi}=1$ TeV and $V_{H}\simeq V_{\textrm{SM}}=246$
GeV we can notice that $\epsilon\lesssim3.8\times10^{-4}$ in order
to satisfy the limit coming from red-giant stars analysis. It is straightforward
to show that the smallness of $\epsilon$ is technically natural,
since doing $\epsilon\rightarrow0$ increases the symmetry of the
total Lagrangian. 

The charged sector can also be found analytically. Besides the charged
Nambu-Goldstone eaten by the $W^{\pm}$ gauge boson, the model has
one charged scalar, $C^{\pm}$. It can be written as $C^{\pm}=\frac{1}{\sqrt{V_{H}^{2}+V_{\phi}^{2}\epsilon^{2}}}\left(V_{\phi}\epsilon\, H^{\pm}+V_{H}\,\Phi^{\pm}\right),$
with squared mass given by $m_{C^{\pm}}^{2}=\frac{\kappa_{H\Phi X}V_{H}}{\sqrt{2}}\frac{1}{\epsilon}+\frac{\kappa'_{H\Phi}V_{H}^{2}}{2}+\frac{\kappa_{H\Phi X}V_{\phi}^{2}}{\sqrt{2}V_{H}}\epsilon+\frac{1}{2}\kappa'_{H\Phi}V_{\phi}^{2}\epsilon^{2}$.
Note that when $\epsilon\rightarrow0$ we have that in general $m_{C^{\pm}}\rightarrow\infty$.
However, when this happens, the minimization conditions in Appendix
\ref{Appendix} require that $\kappa_{H\Phi X}\propto\epsilon$. Thus,
$m_{C^{\pm}}$ remains finite.

In order to find the rest of the mass eigenvalues and eigenstates
of the scalar potential (the $CP-$even, $CP-$odd scalars), in general,
we numerically proceed choosing the set of the parameters to satisfy
simultaneously the minimization conditions given in Eqs. (\ref{v1}-\ref{v6}),
the positivity of the squared masses, and the lower boundedness of
the scalar potential. All these constraints are always checked numerically.
Furthermore, we restrict ourselves to a relevant set of parameters
that allows us to study the dark matter properties in some interesting
cases. Our initial assumptions are: (i) For the sake of simplicity:
$V_{\phi_{1}}=V_{\phi_{2}}=V_{\phi_{3}}=V_{\phi_{X}}\equiv V_{\phi}$
(we have already used this in Eqs. (\ref{majoron}) and in the $C^{\pm}$
charged scalar), $\kappa_{H\Phi}=\kappa'_{H\Phi}=\kappa_{H1}=$$\ \kappa_{H3}=\kappa_{HX}=\kappa_{\Phi1}=\kappa_{\Phi2}=$$\ \kappa_{\Phi3}=\kappa_{\Phi X}=\kappa_{12}=$$\ \kappa_{13}=\kappa_{1X}=\kappa_{23}=$
$\ \kappa_{2X}=\kappa_{3X}=0$ and $\kappa_{123X}=\kappa'_{3X}=\kappa_{123}$
(ii) In order to have the heaviest $CP-$even scalars with similar
masses, we choose: $\lambda_{1}=\lambda_{2}=\lambda_{3}=\lambda_{X}\equiv\lambda_{\phi}$,
and (iii) Due to the stability of the minima, we obtain: $\kappa_{H\Phi X}=V_{\phi}\epsilon$
(see Eq. (\ref{v2})) and $\mu_{H}^{2}=\lambda_{H}V_{H}+\frac{\kappa_{H2}V_{\phi}^{2}}{2}-\frac{V_{\phi}^{3}}{\sqrt{2}V_{H}}\epsilon^{2}$,
$\mu_{\Phi}^{2}=-\frac{V_{H}V_{\phi}}{\sqrt{2}}+\text{\ensuremath{\lambda_{\Phi}}}V_{\phi}^{2}\epsilon^{2}$,
$\mu_{1}^{2}=(\text{\ensuremath{\kappa_{123}}}+\text{\ensuremath{\lambda_{\phi}}})V_{\phi}^{2}$,
$\mu_{2}^{2}=\frac{\kappa_{H2}V_{H}^{2}}{2}+(\kappa_{123}+\text{\ensuremath{\lambda_{\phi}}})V_{\phi}^{2}$,
$\mu_{3}^{2}=(3\kappa_{123}+\text{\ensuremath{\lambda_{\phi}}})V_{\phi}^{2}$,
and $\mu_{X}^{2}=(\kappa_{123}+\text{\ensuremath{\lambda_{\phi}}})V_{\phi}^{2}-\frac{V_{H}V_{\phi}}{\sqrt{2}}\epsilon^{2}$.
The rest of parameters will be set when required.

In general, the squared mass matrices of the $CP-$odd scalars ($M_{CP-\textrm{odd}}^{2}$)
and the $CP-$even scalars ($M_{CP-\textrm{even}}^{2}$) can be written
in powers of $\epsilon$ up to $\epsilon^{2}$, i.e. $M_{i}^{2}=M_{0,i}^{2}+\epsilon M_{1,i}^{2}+\epsilon^{2}M_{2,i}^{2}$
with $i=CP-\textrm{odd},\, CP-\textrm{even}$. In spite of the smallness
of $\epsilon$ and the assumptions made above, it is a hard task to
obtain exact analytical expressions for the mass eigenvalues and mass
eigenstates of these matrices. These can be found perturbatively in
powers of $\epsilon$, though expressions are usually very long and
no more clarifying. In this section we just provide the leading-order
expression of the scalar masses because these yield a good picture
of their exact behavior. 

In the $CP-$odd sector the model has three scalars, $I_{1},\, I_{2},\, I_{3},$
besides the Majoron $J$ in Eq. (\ref{majoron}) and the two Nambu-Goldstone
eaten by the $Z_{1}$ (it is assumed that $Z_{1}$ is the gauge boson
with mass equal to the $Z$ boson in the SM) and $Z_{2}$ boson. Their
masses are given by $m_{I_{1}}=\frac{\sqrt{V_{H}V_{\phi}}}{\sqrt[4]{2}}$
, $m_{I_{2}}=\sqrt{5-\sqrt{7}}\sqrt{-\kappa_{123}}V_{\phi}$, $m_{I_{3}}=\sqrt{5+\sqrt{7}}\sqrt{-\kappa_{123}}V_{\phi}$.
From the previous expressions we see that $\kappa_{123}<0$ in order
to have all masses belonging to reals. It is also straightforward
to see that $I_{1}=\textrm{Im}\,\Phi^{0}+{\cal O}(\epsilon)$. Additionally,
we find that $I_{2}$ and $I_{3}$ are, at $\epsilon$ order, a linear
combination of the $\textrm{Im}\,\phi_{i}$'s with $i=1,2,3,X$. The
$CP-$even sector is more complicated even in the leading order. In
this sector the model has six different eigenstates, $R_{i}$'s, with
masses given by: $m_{R_{1}}=\sqrt{2\lambda_{H}}V_{H}$, $m_{R_{2}}=\frac{\sqrt{V_{H}V_{\phi}}}{\sqrt[4]{2}}$,
$m_{R_{3}}=\sqrt{2\lambda_{\phi}-3.58\left|\kappa_{123}\right|}V_{\phi}$,
$m_{R_{4}}=\sqrt{2\lambda_{\phi}+1.15\left|\kappa_{123}\right|}V_{\phi}$,
$m_{R_{5}}=\sqrt{2\left(\lambda_{\phi}+\left|\kappa_{123}\right|\right)}V_{\phi}$,
$m_{R_{6}}=\sqrt{2\lambda_{\phi}+2.42\left|\kappa_{123}\right|}V_{\phi}$.
$R_{1}$ (which is $\textrm{Re}\, H^{0}+{\cal O}(\epsilon)$) is the
scalar that plays the role of the Higgs scalar boson in this model,
since it couples at tree level to all fermions, giving mass to them
when it gains a VEV, $V_{H}$. Thus, we set its mass equal to $125$
GeV. We find that $\lambda_{H}\simeq0.13-0.14$ gives the correct
value for the Higgs mass. $R_{2}$ is $\textrm{Re}\,\Phi^{0}+{\cal O}(\epsilon)$.
The rest of fields are, in general, combinations of the $\textrm{Re}\,\phi_{i}$'s
with $i=1,\,2,\,3,\, X$. Note that the remaining four $CP-$even
scalars have masses proportional to $V_{\phi}$ and there is not a
criterium to determine precisely their masses. However, we have to
choose the parameters in the scalar potential such that all $m_{R_{i}}$
masses are larger than the $Z_{1}$ boson mass ($m_{Z_{1}}^{2}\approx\frac{g^{2}\left(V_{H}^{2}+V_{\Phi}^{2}\right)}{4\cos^{2}\theta_{W}}=\frac{m_{W}^{2}}{\cos^{2}\theta_{W}}$)
due to the $Z_{1}$ invisible decay width. In other words, if some
of $m_{R_{i}}$ were $<M_{Z_{1}}$ then the $Z_{1}$ boson could decay
through the process $Z_{1}\rightarrow R_{i}+J\rightarrow J+J+J$,
which would contribute to the $Z_{1}$ boson decay width as half of
the decay $Z_{1}\rightarrow\bar{\nu}\nu$ \citep{GonzalezGarcia:1989zh}.
According to the experimental data there is no room for such an extra
contribution \citep{Olive:2014PDG}. 

All expressions above for masses and eigenstates are very useful to
have a general view of the scalar spectrum. However, it is necessary
to work with more precision when calculations of the DM sector are
involved. Thus, from here on, we always work numerically to diagonalize
the squared-mass matrices for both the $CP-$odd and the $CP-$even
scalars. 

Finally, a further comment regarding the $J$ presence is necessary.
Since the Majoron $J$ is massless, it contributes to the density
of radiation in the Universe which is usually parameterized by the
effective neutrino number $N_{\textrm{eff}}$. This parameter specifies
the energy density of relativistic species in terms of the neutrino
temperature. Planck together with WMAP9 polarization data, high-$l$
experiments and the BAO data (Planck + WP + highL + BAO) gives $N_{\textrm{eff}}=3.30_{-0.51}^{+0.54}$
\citep{PLANCK_2013}. In the case that the Majoron $J$ goes out of
equilibrium when the only massive particles left are electrons and
positrons it makes a contribution to $N_{\textrm{eff}}$ equal to
$4/7$ which is in agreement with the current data. In the case when
$J$ decouples in higher temperatures a lower contribution is expected.
For a best treatment see Ref. \citep{Weinberg2013}.

\section{neutrino masses}

\label{sec:Neutrino Masses}The mass Lagrangian for neutrinos, which
comes from Eq. (\ref{lyukawa}) when the neutral scalars gain VEVs,
can be written in matrix form as: 
\begin{equation}
-\lag_{m_{\nu}}=\frac{1}{2}\left[\begin{array}{cc}
\overline{\nu_{L}} & \overline{(n^{c})_{L}}\end{array}\right]\left[\begin{array}{cc}
0 & M_{D}\\
M_{D}^{T} & M_{M}
\end{array}\right]\left[\begin{array}{c}
(\nu^{c})_{R}\\
n_{R}
\end{array}\right]+\textrm{H.c.}\text{,}\label{matriz de massa}
\end{equation}
where $\nu_{L}=\left[\nu_{e}\,\,\nu_{\mu}\,\,\nu_{\tau}\right]_{L}^{T},$
$n_{R}=\left[n_{1}\,\, n_{2}\,\, n_{3}\right]_{R}^{T}\text{.}$ The
Majorana and Dirac mass matrices ($M_{M}$ and $M_{D}$, respectively)
are written as
\begin{equation}
M_{M}=\frac{V_{\phi}}{\sqrt{2}}\left(\begin{array}{ccc}
\mathcal{M}_{11} & \mathcal{M}_{12} & 0\\
\mathcal{M}_{12} & \mathcal{M}_{22} & 0\\
0 & 0 & \mathcal{M}_{33}
\end{array}\right),\qquad M_{D}=\frac{V_{\Phi}}{\sqrt{2}}\left(\begin{array}{ccc}
\mathcal{D}_{11} & \mathcal{D}_{12} & 0\\
\mathcal{D}_{21} & \mathcal{D}_{22} & 0\\
\mathcal{D}_{31} & \mathcal{D}_{32} & 0
\end{array}\right),\label{neutrinomass}
\end{equation}
For $\epsilon\ll1$, i.e. $V_{\Phi}\ll V_{\phi}$, the mass matrix
in Eq. (\ref{matriz de massa}) can be diagonalized using the regular
see-saw mechanism. The masses of the heavy neutrinos, $N_{i}$ with
$i=1,\,2,\,\textrm{DM}$, are related to the energy scale of the VEVs
of the singlets and are given by the eigenvalues of $M_{M}$: $M_{N_{1,2}}=\left[\left(\mathcal{M}_{11}+\mathcal{M}_{22}\right)\mp\sqrt{4\mathcal{M}_{12}^{2}+(\mathcal{M}_{11}-\mathcal{M}_{22})^{2}}\right]V_{\phi}/\left(2\sqrt{2}\right),$
$M_{N_{3}}=\mathcal{M}_{33}V_{\phi}/\sqrt{2}\equiv M_{\textrm{DM}}$.
For simplicity, we set $\mathcal{M}_{12}=0$ and $\mathcal{M}_{11}=\mathcal{M}_{22}$.
Doing so, we have $M_{N_{1,2}}\equiv M_{N}=\mathcal{M}_{11}V_{\phi}/\sqrt{2}$
and $M_{N_{3}}=M_{\textrm{DM}}$. We work with $M_{\textrm{DM}}$
, $M_{N}$ and $V_{\phi}$ as input parameters. Thus, $\mathcal{M}_{11}$
and $\mathcal{M}_{33}$ are expressed in terms of $M_{\textrm{DM}}$
, $M_{N}$ and $V_{\phi}$ as $\mathcal{M}_{11}=\sqrt{2}M_{N}/V_{\phi}$
and $\mathcal{M}_{33}=\sqrt{2}M_{\textrm{DM}}/V_{\phi}$, respectively.

As it is well known, the masses of the light neutrinos, $\nu_{i}$
with $i=1,\,2,\,3$, are given by the eigenvalues of the matrix $M_{\nu}=M_{D}M_{M}^{-1}M_{D}^{T}$.
From Eq. (\ref{neutrinomass}), it can be seen that $\det M_{\nu}=0$.
It implies that at least one of the light neutrino masses is zero.
The minimal requirement for the parameters in $M_{M}$ and $M_{D}$
is that these have to provide the light neutrino masses and mixing
angles consistent with the oscillation neutrino constraints. There
are other constraints on neutrino masses such as $\sum_{i=1}^{3}m_{\nu_{i}}<0.23$
eV coming from Planck collaboration \citep{PLANCK_2013} that we are
going to consider below. 

Now, we proceed analytically making the ansatz that $M_{\nu}$ is
diagonalized by the tri-bimaximal-Cabbibo (TBC) matrix, $U_{\textrm{TBC}}$
\citep{KingLuhn:2013neutrinomixing}, i.e. $U_{\textrm{TBC}}^{T}M_{\nu}U_{\textrm{TBC}}=\hat{M}_{\nu}=\textrm{diag}\left(m_{1},\, m_{2},\, m_{3}\right)$.
For the sake of simplicity, we parametrize $m_{1}=x-y,$ $m_{2}=2x+y$,
$m_{3}=2\nu+x-y$, and work in a basis where the charged lepton mass
matrix is diagonal. $U_{\textrm{TBC}}$ can be written as 
\begin{equation}
U_{\textrm{TBC}}=\left(\begin{array}{ccc}
\sqrt{\frac{2}{3}}\left(1-\frac{\lambda^{2}}{4}\right) & \frac{1}{\sqrt{3}}\left(1-\frac{\lambda^{2}}{4}\right) & \frac{\lambda}{\sqrt{2}}\\
-\frac{1}{\sqrt{6}}\left(1+\lambda\right) & \frac{1}{\sqrt{3}}\left(1-\frac{\lambda}{2}\right) & \frac{1}{\sqrt{2}}\left(1-\frac{\lambda^{2}}{4}\right)\\
\frac{1}{\sqrt{6}}\left(1-\lambda\right) & -\frac{1}{\sqrt{3}}\left(1+\frac{\lambda}{2}\right) & \frac{1}{\sqrt{2}}\left(1-\frac{\lambda^{2}}{4}\right)
\end{array}\right)+{\cal O}(\lambda^{3}),\label{TBC}
\end{equation}
where we have chosen $\delta=0$ (see \citep{KingLuhn:2013neutrinomixing}).
$U_{\textrm{TBC}}$ leads to the mixings: $\sin\theta_{12}=\frac{1}{\sqrt{3}}$,
$\sin\theta_{13}=\frac{\lambda}{\sqrt{2}}$ and $\sin\theta_{23}=\frac{1}{\sqrt{2}}$.
We choose $\lambda=0.218174$ to be consistent with the experimental
limits \citep{Olive:2014PDG} for neutrinos. It is remarkable that
this $\lambda$ value is consistent with the relationship $\sin\theta_{C}\approx\lambda$
where is the Wolfenstein parameter. In order to set the parameters
in $M_{M}$ and $M_{D}$, we match $U_{\textrm{TBC}}\hat{M}_{\nu}U_{\textrm{TBC}}^{T}$
and $M_{D}M_{M}^{-1}M_{D}^{T}$ as it should be if our ansatz is supposed
to work. Note that we have used $U_{\textrm{TBC}}U_{\textrm{TBC}}^{T}=1+{\cal O}(\lambda^{4})$
and $U_{\textrm{TBC}}^{T}U_{\textrm{TBC}}=1+{\cal O}(\lambda^{4})$.
In addition, we have one more degree of freedom to choose because
the neutrino mass hierarchy is yet unknown. The neutrino mass hierarchy
can be either normal ($m_{1}<m_{2}<m_{3}$) or inverted ($m_{3}<m_{1}<m_{2}$).
We separately consider them.

In the case of normal hierarchy we choose $m_{1}=0$ since $\det\, M_{\nu}=0$.
Doing so, $m_{2}=3y$ and $m_{3}=2\nu$. Hence, we find
\begin{equation}
U_{\textrm{TBC}}\hat{M}_{\nu}U_{\textrm{TBC}}^{T}=\frac{1}{16}\begin{array}[t]{c}
\left(\begin{array}{cc}
y\left(-4+\lambda^{2}\right)+\lambda^{2}\nu & 2\left(-4+\lambda^{2}\right)\left(y\left(-2+\lambda\right)-2\lambda\nu\right)\\*
* & \left(-2+\lambda\right)^{2}\left(4y+\left(2+\lambda\right)^{2}\nu\right)\\*
* & *
\end{array}\right.\\
\qquad\qquad\qquad\qquad\left.\begin{array}{c}
2\left(-4+\lambda^{2}\right)\left(y\left(2+\lambda\right)-2\lambda\nu\right)\\
\left(-4+\lambda^{2}\right)\left(4y+\left(-4+\lambda^{2}\right)\nu\right)\\
\left(2+\lambda\right)^{2}\left(4y+\left(-2+\lambda\right)^{2}\nu\right)
\end{array}\right),
\end{array}\label{MTB-3}
\end{equation}
where ``$*$'' means that the assigned matrix element is equal to
its transpose element. The matrix $M_{D}M_{M}^{-1}M_{D}^{T}$ is written
as
\begin{eqnarray}
M_{D}M_{M}^{-1}M_{D}^{T} & = & K\left(\begin{array}{ccc}
\mathcal{D}_{11}^{2}+\mathcal{D}_{12}^{2} & \mathcal{D}_{11}\mathcal{D}_{21}+\mathcal{D}_{12}\mathcal{D}_{22} & \mathcal{D}_{11}\mathcal{D}_{31}+\mathcal{D}_{12}\mathcal{D}_{32}\\
* & \mathcal{D}_{21}^{2}+\mathcal{D}_{22}^{2} & \mathcal{D}_{21}\mathcal{D}_{31}+\mathcal{D}_{22}\mathcal{D}_{32}\\
* & * & \mathcal{D}_{31}^{2}+\mathcal{D}_{32}^{2}
\end{array}\right).\label{lightneutrinos}
\end{eqnarray}
where we have defined the dimensional constant $K\equiv\frac{V_{\Phi}^{2}}{\sqrt{2}\mathcal{M}_{11}V_{\phi}}=\frac{V_{\phi}^{2}\epsilon^{2}}{2M_{N}}$.
Matching Eq. (\ref{MTB-3}) to Eq. (\ref{lightneutrinos}) we have
a system of six independent equations. We were not able to solve analytically
that system for the six general variables $\mathcal{D}_{11}$, $\mathcal{D}_{12}$,
$\mathcal{D}_{21}$, $\mathcal{D}_{22}$, $\mathcal{D}_{31}$, $\mathcal{D}_{32}$.
However, if we set $\mathcal{D}_{12}=0$ \hyperlink{label}{\footnote[1]{We have numerically solved the equations finding always ${\cal{D}}_{12}=0$, what justifies our choice.}}\hypertarget{label}{},
we can solve it analytically for the remaining five variables and
obtain the following four solutions:
\begin{eqnarray}
\mathcal{D}_{11} & = & s_{1}\times\frac{\sqrt{r}}{4\sqrt{K}},\label{d11}\\
\mathcal{D}_{21} & = & s_{1}\times\frac{\left(\lambda^{2}-4\right)((\lambda-2)y-2\lambda\nu)}{2\sqrt{Kr}},\\
\mathcal{D}_{22} & = & s_{2}\times\frac{\sqrt{\left(\lambda^{4}-16\lambda+16\right)^{2}\nu y^{2}}}{4\sqrt{Kry}},\\
\mathcal{D}_{31} & = & s_{1}\times\frac{\left(\lambda^{2}-4\right)((\text{\ensuremath{\lambda}}+2)y-2\lambda\nu)}{2\sqrt{Kr}},\\
\mathcal{D}_{32} & = & s_{2}\times\frac{\left(\lambda^{4}+16\lambda+16\right)\sqrt{\left(\text{\ensuremath{\lambda}}^{4}-16\lambda+16\right)\nu y^{2}}}{4\sqrt{Kry}\left(\text{\ensuremath{\lambda}}^{4}-16\text{\ensuremath{\lambda}}+16\right)},\label{d32}
\end{eqnarray}
where $s_{1}\equiv\left\{ -1,-1,+1,+1\right\} $, $s_{2}\equiv\left\{ -1,+1,-1,+1\right\} $
and $r\equiv16\lambda^{2}\nu+\left(\lambda^{2}-4\right)^{2}y$.

Now that we have the solutions for the $\mathcal{D}_{ij}$ in terms
of $y$, $\nu$ and $K$, let's find $y$ and $\nu$ from $\Delta m_{\textrm{sun}}^{2}=m_{2}^{2}-m_{1}^{2}=9y^{2}$
and $|\Delta m_{\textrm{atm}}^{2}|=m_{3}^{2}-m_{1}^{2}=4\nu^{2},$
with $\Delta m_{\textrm{sun}}^{2}=7.53\times10^{-5}$ $\textrm{eV}^{2}$
and $|\Delta m_{\textrm{atm}}^{2}|=2.52\times10^{-3}$ $\textrm{eV}^{2}$
\citep{Olive:2014PDG}. We solve these equations obtaining $y\approx2.89252\times10^{-3}$
eV and $\nu\approx2.50998\times10^{-2}$ eV. Neither $y$ nor $\nu$
may be negative because $m_{2}$ and $m_{3}$ are positive. We then
find the masses values: $m_{1}=0,\, m_{2}\approx8.677556\times10^{-3}\textrm{ eV},\, m_{3}\approx5.01996\times10^{-2}\textrm{ eV}$
which shows a normal mass hierarchy. 

In order to determine completely the $\mathcal{D}_{ij}$ values, we
still have to find $K$. From Eq. (\ref{lightneutrinos}) and assuming
${\cal O}(\mathcal{D}_{ij})\sim1$, we have that $K\lesssim10^{-10}$
GeV sets the neutrino masses in sub-eV mass scale. However, the value
of $K$ can not be taken arbitrarily small due to one-loop induced
processes violating lepton flavor (LFV). Specifically, we consider
LFV processes such as $l_{i}\rightarrow l_{j}+\gamma$, where $i=\mu,\,\tau$
and $j=e,\,\mu$, respectively. This model has one-loop contributions
to these kinds of processes since charged leptons couple to charged
scalars and right-handed heavy neutrinos. The branching ratio is estimated
as $\textrm{Br}\left(l_{i}\rightarrow l_{j}+\gamma\right)=\frac{96\pi^{3}\alpha}{G_{F}^{2}m_{l_{i}}^{4}}\left(\left|f_{M1}\right|^{2}+\left|f_{E1}\right|^{2}\right)$
\citep{Branching_ratio_theory}, where $\alpha\simeq1/137$, $G_{F}\simeq1.16\times10^{-5}$
GeV$^{-2}$ is the Fermi constant and $f_{M1}=f_{E1}=\sum_{k=1}^{3}\frac{{\cal D}_{ik}{\cal D}_{jk}}{4\left(4\pi\right)^{2}}\frac{m_{l_{i}}^{2}}{m_{C^{\pm}}^{2}}F_{2}\left(\frac{M_{N_{k}}^{2}}{m_{C^{\pm}}^{2}}\right)$
with $F_{2}\left(x\right)=\frac{1-6x+3x^{2}+2x^{3}-6x^{2}\ln x}{6\left(1-x\right)^{4}}$.
The present upper bounds for $\textrm{Br}\left(\mu\rightarrow e+\gamma\right)$
and $\textrm{Br}\left(\tau\rightarrow\mu+\gamma\right)$ are $<5.7\times10^{-13}$
and $<4.4\times10^{-8}$ \citep{Olive:2014PDG}, respectively. The
dependence of the $\textrm{Br}\left(l_{i}\rightarrow l_{j}+\gamma\right)$
on $K$ value arises through $f_{M1}$ and $f_{E1}$ which depend
on ${\cal D}_{ij}$ values (see Eqs. (\ref{d11}-\ref{d32})). Also,
note that $\textrm{Br}\left(l_{i}\rightarrow l_{j}+\gamma\right)$
weakly depends on $M_{N_{k}}$ because $F_{2}\left(x\right)$ does
not drastically depend on its argument. Thus, we can find a lower
bound for $K$ imposing the experimental upper bounds $\textrm{Br}\left(\mu\rightarrow e+\gamma\right)$
and $\textrm{Br}\left(\tau\rightarrow\mu+\gamma\right)$. We find
that for $200\textrm{ GeV}<M_{N_{1,2}}<1000\textrm{ GeV}$, the $K$
value has to be $>2.72\times10^{-9}$ GeV. We have also used $m_{C^{\pm}}=416.986$
GeV, which is the value used in the DM analysis and is its correct
limit at ${\cal O}(\epsilon)$. Now, we can finally find the values
for $\mathcal{D}_{ij}$. For instance, for the first solution displayed
in Eqs. (\ref{d11}-\ref{d32}), we have $\mathcal{D}_{11}\approx-0.0754,$
$\mathcal{D}_{21}\approx-0.1492,$ $\mathcal{D}_{22}\approx-0.125\text{,}$
$\mathcal{D}_{31}\approx-0.042$ and $\mathcal{D}_{32}\approx-0.1947$.
It is important to say that the lower limit on $K$ imposes a constraint
on $\epsilon\gtrsim1.97\times10^{-6}$ (we have fixed $M_{N_{1,2}}=1000/\sqrt{2}$
GeV and $V_{\phi}$=1 TeV). This constraint on $\epsilon$ and the
one coming from the safety of the Majoron $J$ (see Section \ref{sec:Scalar Sector})
imply that $1.97\times10^{-6}\lesssim\epsilon\lesssim3.8\times10^{-4}$. 

Experiments on $0\nu\beta\beta$ constrain the effective Majorana
mass $\left|m_{ee}\right|=\left|c_{13}^{2}\left(m_{1}c_{12}^{2}e^{i\delta_{1}}+m_{2}s_{12}^{2}e^{i\delta_{2}}\right)+m_{3}e^{2i\phi_{CP}}s_{13}^{2}\right|$
and the strongest one is up to now $\left|m_{ee}\right|<0.27-0.65\textrm{ eV},$
$90\%$ C. L. \citep{mee_future,mee_CUORE_2015}. We are not interested
in CP violation nor phases in the leptonic mixing matrix, therefore
we end up finding $m_{ee}=0.00376994$ eV. Also, data coming from
Planck collaboration \citep{PLANCK_2013} constrain the sum of the
light neutrinos to be lower than $<0.23\textrm{ eV}$. Clearly, it
is satisfied since $m_{1}+m_{2}+m_{3}\thicksim\mathcal{O}\left(10^{-2}\right)\,$
eV.

For the inverted mass hierarchy, the procedure is very similar to
the one shown above, thus, we present only the main results. In this
case, we have $m_{3}=0$, as opposed to $m_{1}=0$. It yields the
parametrization $m_{1}=x-y$, $m_{2}=2x+y$ and $m_{3}=0$. For the
$\mathcal{D}_{ij}$ values, we have $\mathcal{D}_{12}=0$ and
\begin{eqnarray}
\mathcal{D}_{11}=s_{1}\times\frac{\left(\lambda^{2}-4\right)\sqrt{x}}{4\sqrt{K}}, &  & \mathcal{D}_{21}=s_{1}\times\frac{\lambda x-2y}{2\sqrt{Kx}},\label{d11i}\\
\mathcal{D}_{22}=-\mathcal{D}_{32}=s_{2}\times\frac{\sqrt{(x-y)(x+2y)}}{\sqrt{2Kx}}\text{,} &  & \mathcal{D}_{31}=s_{1}\times\frac{\text{\ensuremath{\lambda}}x+2y}{2\sqrt{Kx}}\text{,}\label{d32i}
\end{eqnarray}
where $s_{1}$, $s_{2}$ and $K$ are defined as above. The values
for $x$ and $y$ are found from $\Delta m_{\textrm{sun}}^{2}=m_{2}^{2}-m_{1}^{2}=3y(2x+y)>0,$
and $|\Delta m_{\textrm{atm}}^{2}|=m_{3}^{2}-m_{1}^{2}=(x-y)^{2}.$
If $x>y$, we have $3y(2x+y)=7.53\times10^{-5}$ $\text{eV}{}^{2}$
and $x-y=(2.52\times10^{-3})^{1/2}$ eV. Solving for $x$ and $y$,
we have $x\approx5.04478\times10^{-2}$ eV and $y\approx2.48162\times10^{-4}$
eV (there is another solution: $x\approx1.64850\times10^{-2}$ eV
and $y\approx-3.37146\times10^{-2}$ eV, but we choose the first one
to work with). We can then find the mass values: $m_{1}\approx5.01996\times10^{-2}\textrm{ eV},\, m_{2}\approx5.09441\times10^{-2}\textrm{ eV},\, m_{3}=0$
which show an inverted mass hierarchy, as said in the beginning. The
LFV processes require that $K\gtrsim1.77\times10^{-9}$ GeV. Using
these values for $K$, $x$, $y$ and $\lambda$, we obtain, for the
first solution displayed in Eqs. (\ref{d11i}-\ref{d32i}), $\mathcal{D}_{11}\approx0.0523$,
$\mathcal{D}_{12}\approx0,$ $\mathcal{D}_{21}\approx-0.0056,$ $\mathcal{D}_{22}\approx-0.0379$,
$\mathcal{D}_{31}\approx-0.0061$ and $\mathcal{D}_{32}\approx0.0379$.
Similarly to the normal hierarchy, we have that $1.58\times10^{-6}\lesssim\epsilon\lesssim3.8\times10^{-4}$
in order to satisfy the experimental bounds for $\textrm{Br}\left(\mu\rightarrow e+\gamma\right)$
and $\textrm{Br}\left(\tau\rightarrow\mu+\gamma\right)$. 

For the $m_{ee}$ limit, we find $m_{ee}=0.0492258$ eV, which is
below the latest experimental limit \citep{mee_future,mee_CUORE_2015}.
Regarding the Planck limit \citep{PLANCK_2013}, it is satisfied because
$m_{1}+m_{2}+m_{3}\simeq0.101\,$ eV.

\section{dark matter}

\label{sec:dark_matter}As previously mentioned, this model has an
almost automatic $\ZZ_{2}$ symmetry acting on $n_{R3}$, i.e. $\ZZ_{2}\left(n_{R3}\right)=-n_{R3}$.
We have imposed it to be exact in the total Lagrangian by removing
just one term. Thus, $n_{R3}$ is stable and it can, in principle,
be a DM candidate. From here on, we consider $N_{\textrm{DM}}$ (which
is equal to $n_{R3}$, the difference being that $N_{\textrm{DM}}$
is a mass basis field and the former a symmetry basis one) as a DM
candidate and verify whether it satisfies the current experimental
data. These data come essentially from investigations of Planck collabo\textcolor{black}{ration
\citep{PLANCK_2013} wh}ich constrains the DM relic density to be
$\Omega_{\textrm{DM}}h^{2}=0.1193\pm0.0014$; and from direct detection
(DD) limits of LUX \citep{LUX_2014}, XENON100 \citep{XENON100_2012}
and SuperCDMS \citep{SuperCDMS_2014}, which constrain the cross section,
for scattering off nucleon, to be smaller than $7.6\times10^{-10}\textrm{ pb}$
for WIMP mass of $33$ GeV. We will consider these constraints below.

\subsection{Relic Abundance}

In order to find the present DM relic density, $\Omega_{\textrm{DM}}h^{2}$,
coming from the $N_{\textrm{DM}}$ Majorana fermion, we must solve
the Boltzmann differential equation. This standard procedure is well
d\textcolor{black}{escribed in Refs. \citep{Gondolo_Gelmini,Griest_Seckel}.
Here, we a}re not going to enter in its details because we have used
the packages ${\tt Feynrules}$ \citep{FeynRules_2014}, ${\tt Calchep}$
\citep{CalcHEP_2013} and ${\tt MicrOMEGAs}$ \citep{MicrOMEGAs_2015}.
The first two being auxiliary to the third that calculates $\Omega_{\textrm{DM}}h^{2}$
for a given model which contains WIMPs.

\begin{figure}
\includegraphics[scale=0.7]{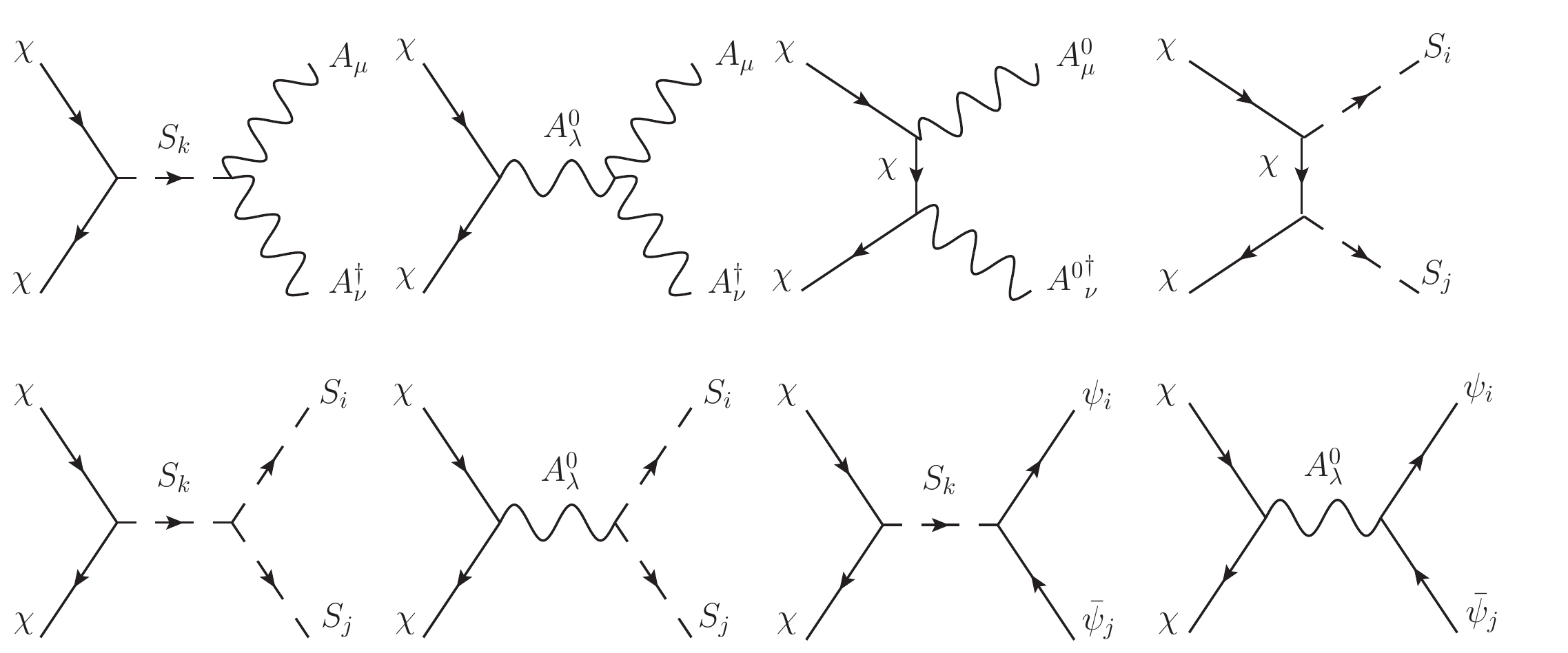}

\protect\caption{Feynman diagrams which represent the main annihilation processes that
contribute to the calculation of $\Omega_{\textrm{DM}}h^{2}$. We
have defined: $\chi\equiv N_{\textrm{DM}}$; $S_{i}\equiv R_{i},\, I_{i}$;
$A_{\mu}\equiv Z_{1\mu},\, W_{\mu}^{\pm}$; and $A_{\mu}^{0}\equiv Z_{1\mu}$.\label{Feynman_diagrams}}
\end{figure}

In Fig. (\ref{Feynman_diagrams}), we show the processes which mainly
contribute to the DM annihilation cross section, and so lead to the
present relic density. All of them depend on the parameters in the
Lagrangians given in Eqs. (\ref{lyukawa}), (\ref{potential}) and
on the kinetic terms involving the covariant derivatives. We have
already fixed most of those parameters in Secs. \ref{sec:Scalar Sector}
and \ref{sec:Neutrino Masses}. However, $g$, $g_{Y^{\prime}}$,
$g_{B-L}$, $\lambda_{H}$, $\lambda_{\phi}$, $\lambda_{\Phi}$,
$\kappa_{H\textrm{2}}$, $\kappa_{\textrm{123}}$ and $M_{\textrm{DM}}$
remain still free. The first three parameters $g$, $g_{Y^{\prime}}$,
$g_{B-L}$ are the gauge coupling constants of the SU$(2)_L$, $\textrm{U}\left(1\right)_{Y^{'}}$
and $\textrm{U}\ensuremath{(1)_{{\cal {B}-{\cal {L}}}}}$ groups,
respectively. Roughly speaking, these couplings and the VEVs together
determine the masses of the gauge bosons. The VEVs have already been
set in the previous sections. In addition, $g$ can be set equal to
$0.652$ due to the $W^{\pm}$ mass. $g_{Y^{\prime}}$ and $g_{B-L}$
mainly determine the masses of the $Z_{2}$ gauge boson and its mixing
with $Z_{1}$ in the neutral current. From precision electroweak studies
\citep{beta_1,beta_2,beta_3}, its mixing, given by $\tan\beta$,
has to be $\lesssim10^{-3}$ (see Ref. \citep{Bruce:2014DM} for an
analytical expression of $\tan\beta$). Furthermore, $M_{Z_{2}}/g_{B-L}\gtrsim6\,\textrm{TeV}$
\citep{MZ2_1,MZ2_2}. We find that working with $g_{Y^{\prime}}=0.506$
and $g_{B-L}=0.505$, we obtain $\tan\beta\simeq2\times10^{-4}$ and
$M_{Z_{2}}\simeq4.7\,\textrm{TeV}$, as well as the known SM gauge
bosons masses. 

Now, the $\lambda_{H}$ parameter is chosen to be $0.13\leq\lambda_{H}\leq0.14$
because it is the main responsible for the Higgs mass, $M_{\textrm{Higgs}}=125\,\textrm{GeV}$,
when $\kappa_{H\textrm{2}}\leq0.1$. In principle, the $\lambda_{\Phi}$
and $\lambda_{\phi}$ parameters can take a wide range of values.
Here, we have set $\lambda_{\Phi}=0.5$ and $\lambda_{\phi}=0.8$,
and thus we have the non-SM scalar masses larger than the SM particle
masses. The $\kappa_{H\textrm{2}}$, $\kappa_{\textrm{123}}$ and
$M_{\textrm{DM}}$ parameters have been scanned in a broad region
of values. Specifically, we have iterated the ${\tt MicrOMEGAs}$
package for the DM mass within the range $10\textrm{ GeV}\leq M_{\textrm{DM}}\leq1000\textrm{ GeV}$,
taking into account different values of $\kappa_{H\textrm{2}}$ and
$\kappa_{\textrm{123}}$, and leaving the remaining parameters constant.
In general, we have worked with $\kappa_{H\textrm{2}}=0.1,\,10^{-2},\,10^{-4}$
and $-0.56\times\lambda_{\phi}\lesssim\kappa_{\textrm{123}}\leqslant0$
($0.56\times0.8=0.448$). The last choice because we must assure that
all the scalar masses are real (we obtain a slightly more constraining
condition on $\kappa_{\textrm{123}}$ if we impose that all $CP-$even
scalar must have masses larger than the Higgs boson, i.e. $\frac{1}{3.58V_{\phi}^{2}}\left(m_{\textrm{Higgs}}^{2}-2\lambda_{\phi}V_{\phi}^{2}\right)\approx-0.442<\kappa_{\textrm{123}}\leqslant0$).
Also, it is important to note that $\kappa_{\textrm{123}}$ controls
the scalar trilinear vertices between scalars. 

Regarding the $\kappa_{H\textrm{2}}$ parameter, we find that, in
our scenario, it largely governs the invisible Higgs width $\Gamma_{\textrm{Higgs}}^{\textrm{Inv }}$
to non-SM particles. It is because $\kappa_{H\textrm{2}}$ induces
mixing between $\textrm{Re}\, H^{0}$ and $\textrm{Re}\,\phi_{2}$
and thus it mostly determines the coupling Higgs-$JJ$, $C_{hJJ}$,
since $J$ has a component in $\textrm{Im}\,\phi_{2}$. This $C_{hJJ}$
coupling induces a tree-level contribution to the $\Gamma_{\textrm{Higgs}}^{\textrm{Inv }}$
given by $C_{hJJ}^{2}/32\pi m_{\textrm{Higgs}}$. Under the assumption
that the production and decays of the Higgs are correctly described
by the SM aside perhaps from decay into new unobserved particles,
the branching ratio for the Higgs decay into new invisible particles,
$\textrm{Br}{}_{\textrm{Higgs}}^{\textrm{Inv}}$, is known to be $\lesssim10\%-15\%$
\citep{Ellis_2013,Giardino2014,Atlascollaboration,CMScollaboration}.
As $\kappa_{H\textrm{2}}<0.2$ we find that the $\textrm{Br}{}_{\textrm{Higgs}}^{\textrm{Inv}}$
remains under this value for $-0.442\lesssim\kappa_{\textrm{123}}\leqslant0$.
We have been conservative choosing $\kappa_{H\textrm{2}}\leq0.1$
for all results.

\begin{figure}
\includegraphics[scale=0.4]{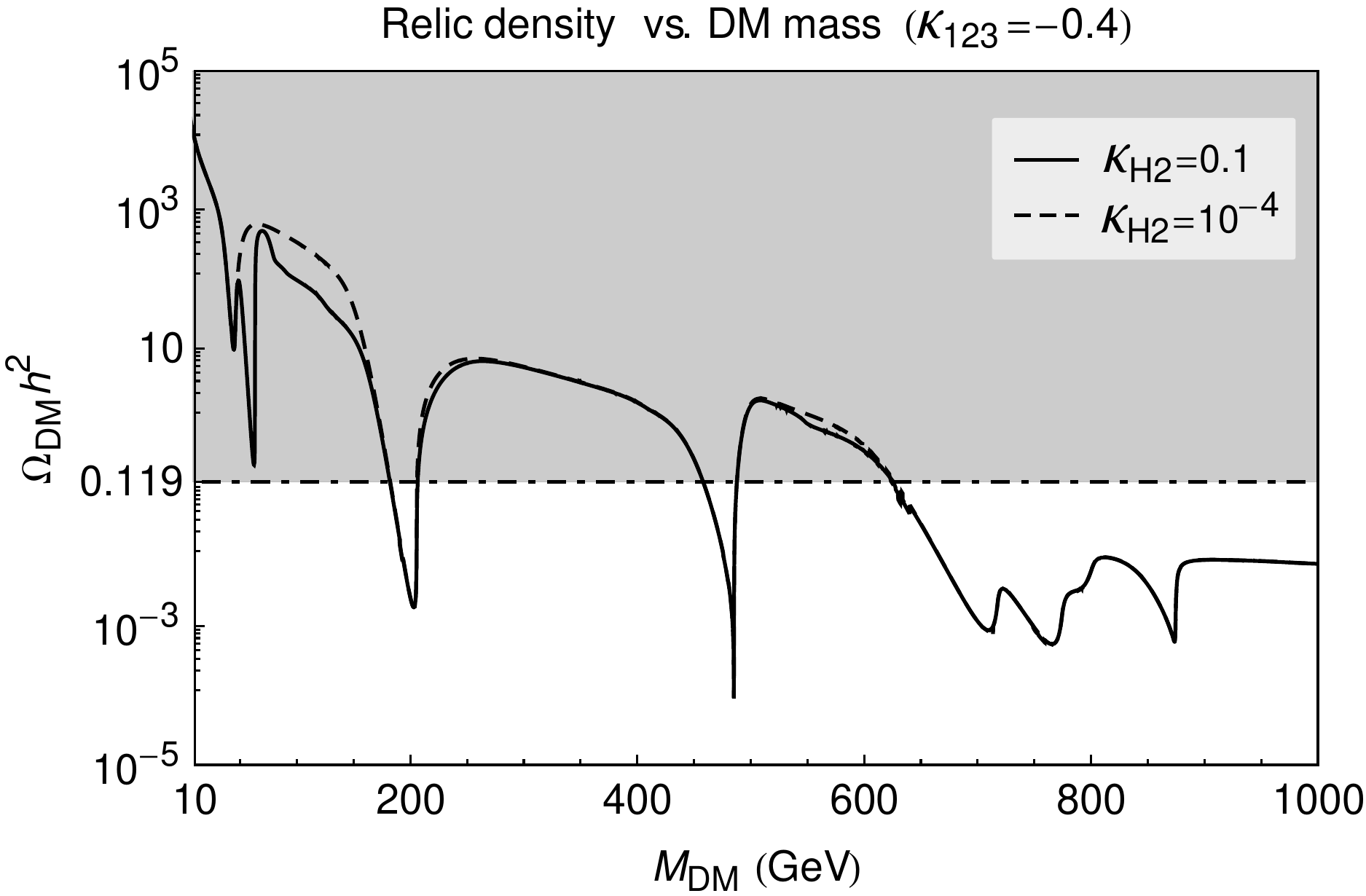}\includegraphics[scale=0.4]{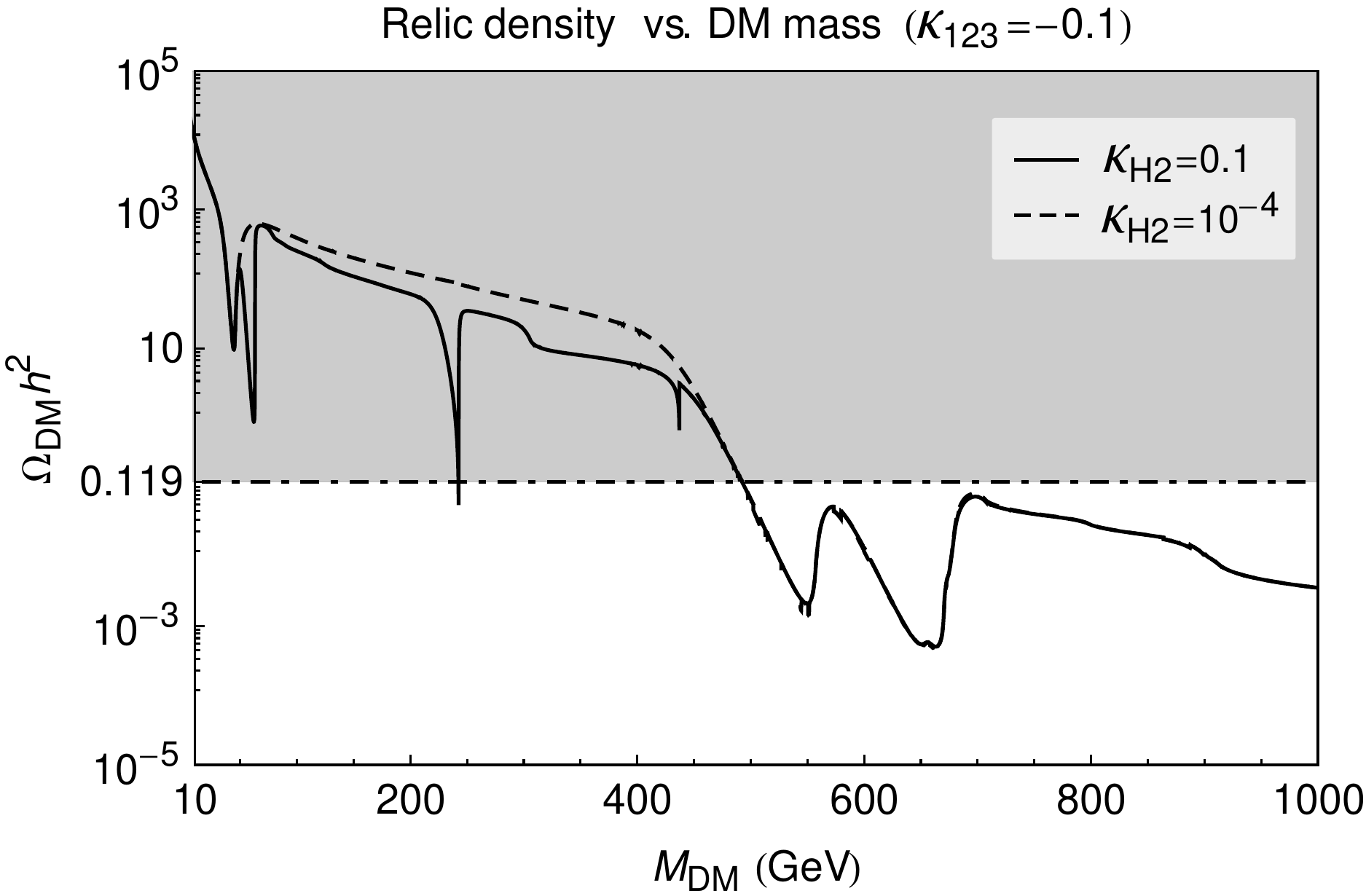}

\protect\caption{Figures displaying the dependence of $\Omega_{\textrm{DM}}h^{2}$
on $M_{\textrm{DM}}$, for $\kappa_{\textrm{123}}=-0.4$ (the figure
on the left) and $\kappa_{\textrm{123}}=-0.1$ (the figure on the
right). Each figure shows two cases corresponding to $\kappa_{H\textrm{2}}=10^{-1},10^{-4}$.
The dot-dashed line is the Planck $\Omega_{\textrm{DM}}h^{2}$. The
gray region means the DM candidate would be excluded, and the white
one means that the DM candidate can still make some part of the DM
content of the Universe\label{fig:Omega1}.}
\end{figure}

Taking into account all aforementioned considerations on the parameters,
we plot, in Fig. \ref{fig:Omega1}, $\Omega_{\textrm{DM}}h^{2}$ versus
$M_{\textrm{DM}}$ for $\kappa_{H\textrm{2}}=10^{-1},\,10^{-4}$,
with $\kappa_{\textrm{123}}=-0.4$ (the figure on the left) and $\kappa_{\textrm{123}}=-0.1$
(the figure on the right), respectively. The region in gray is ruled
out because $\Omega_{\textrm{DM}}h^{2}$ is overabundant. The dot-dashed
line is the $\Omega_{\textrm{DM}}h^{2}=0.1193$ value reported by
Planck. In general, we find that depending on the $M_{\textrm{DM}}$,
various annihilation channels are important and clearly some resonances
are visible. Resonances are, in general, found in $m_{\text{mediator}}/2$.
Thus, for convenience, we give here the scalar masses for both figures
in Fig. (\ref{fig:Omega1}). For the case with $\kappa_{123}=-0.4$
(both values of $\kappa_{H2}$) we approximately have $m_{R_{i}}\simeq125.0,\,417.0,\,411.3,\,1435.8,\,1549.2,\,1603.0$
GeV, $M_{I_{i}}\simeq417.0,\,970.4,\,1748.8$ GeV, and $M_{C^{\pm}}\simeq417.0\,\textrm{GeV}$.
On the other hand, for the case with $\kappa_{\textrm{123}}=-0.1$
(both values of $\kappa_{H2}$), we approximately have $m_{R_{i}}\simeq125.0,\,417.0,\,1114.6,\,1309.7,\,1341.6,\,1357.3$
GeV, $M_{I_{i}}\simeq417.0,\,485.2,\,874.4$ GeV, $M_{\Phi^{\pm}}\simeq417.0\,\textrm{GeV}$.
In all cases we have the Majoron $J$.

\begin{figure}
\includegraphics[scale=0.8]{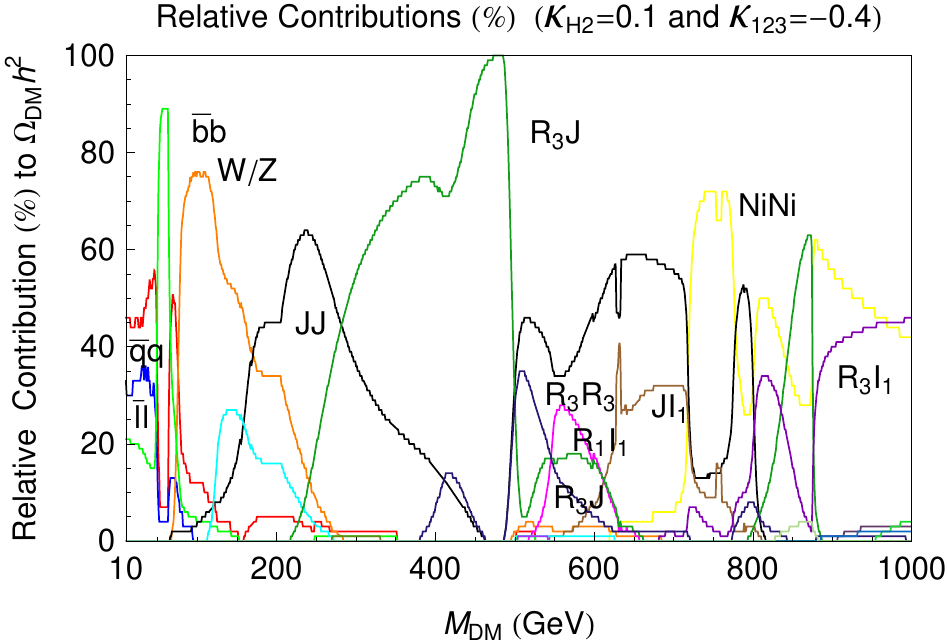}\includegraphics[scale=0.8]{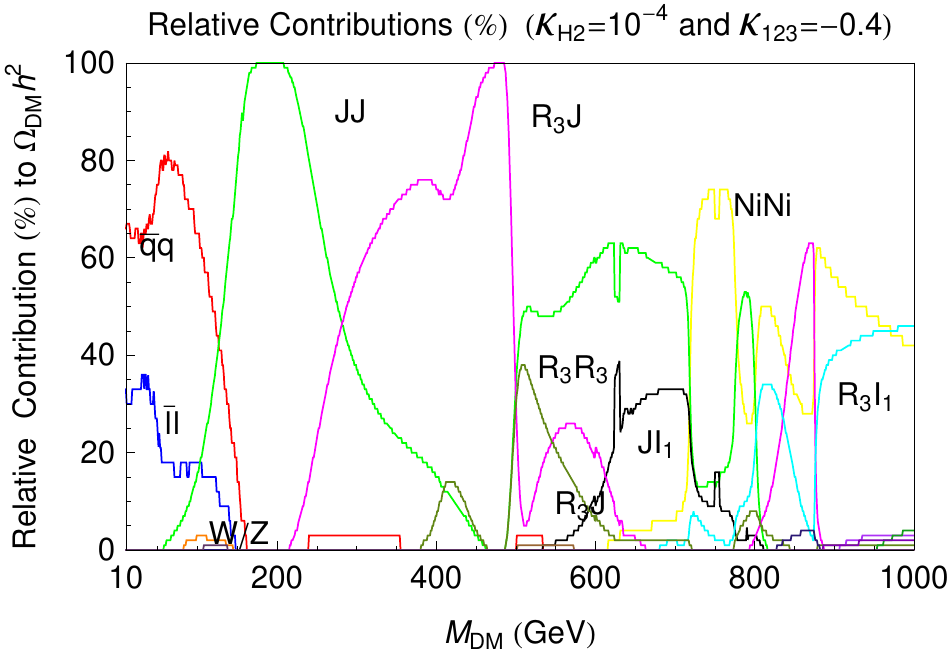}

\includegraphics[scale=0.8]{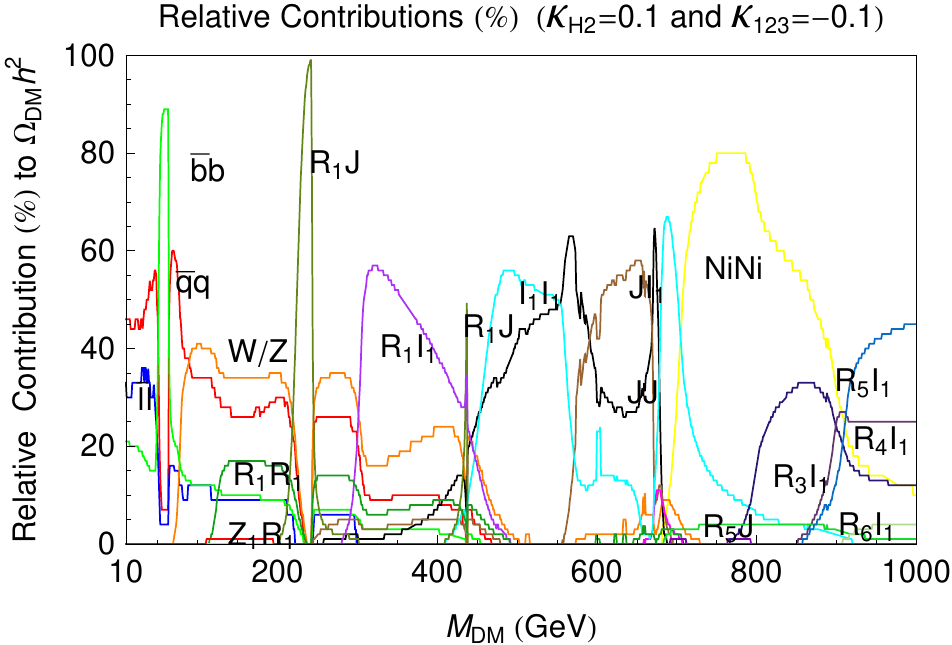}\includegraphics[scale=0.8]{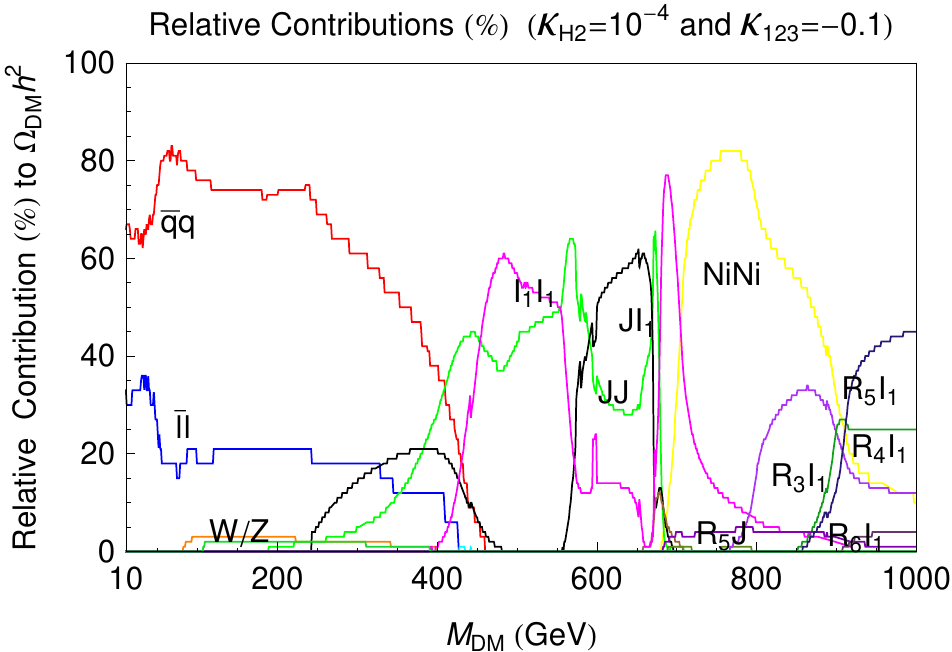}

\protect\caption{(Color online) Figures displaying the relative contributions (\%)
of the main annihilation channels to the DM relic abundance. The cases
with $\kappa_{H\textrm{2}}=10^{-1},\,10^{-4}$ and $\kappa_{\textrm{123}}=-0.1,\,-0.4$
are shown. \label{fig:Omega-contributions}}
\end{figure}

In order to better comprehend the annihilation processes and their
contributions contained in the curves in Fig. \ref{fig:Omega1}, we
plot Fig. \ref{fig:Omega-contributions} which shows the relative
contributions to $\Omega_{\textrm{DM}}h^{2}$ of the main DM annihilation
channels. Let's consider some relevant regions. For $M_{\textrm{DM}}$
less than $80$ GeV we have in general two resonances. The first one
is due to the interchange of the $Z_{1}$ gauge boson in the $s-$channel.
It is located at $M_{\textrm{DM}}=M_{Z_{1}}/2\approx45.6$ GeV and
remains there even when $\kappa_{H2}$ is set to $10^{-4}$. It is
so because it depends on the $N_{\textrm{DM}}-N_{\textrm{DM}}-Z_{1}$
coupling via neutral currents. Since this coupling arises from the
covariant derivatives, it is independent on the $\kappa_{H2}$ value.
In contrast, the second resonance, which arises by the $s-$channel
interchange of the Higgs boson (located in $m_{\text{Higgs}}/2\approx62.5$
GeV), disappears when $\kappa_{H2}=10^{-4}$. This is understood realizing
that $N_{\textrm{DM}}$ couples to the Higgs boson via the term $\frac{1}{2}\frac{\sqrt{2}M_{\textrm{DM}}}{V_{\phi}}\overline{(n_{R3})^{c}}n_{R3}\phi_{2}$
and since Higgs component in $\phi_{2}$ depends on $\kappa_{H2}$,
it is clear that the smaller $\kappa_{H2}$ the smaller $N_{\textrm{DM}}-N_{\textrm{DM}}-$Higgs
coupling.\textcolor{black}{{} In this region of masses we also notice
that $N_{\textrm{DM}}$ annihilation processes into quark-antiquark
pair (in special into $b\bar{b}$ quarks for }$\kappa_{H2}=10^{-1}$\textcolor{black}{)
are the dominant for both figures. These occur via Higgs mediation.
Annihilation processes into neutrinos via the $Z_{1}$ interchange
are also important (}$\sim25\%$\textcolor{black}{). }This is also
true for both values of $\kappa_{H2}$ and for both $\kappa_{\textrm{123}}=-0.4$
and $-0.1$. 

As $M_{\textrm{DM}}$ increases from $80\,\textrm{GeV}$ to $120$
GeV, and as long as $\kappa_{\textrm{123}}=-0.4$ and $\kappa_{H2}=10^{-1}$,
the annihilation into gauge bosons ($W^{\pm}/Z_{1}$) are dominant
(in particular into $W^{+}W^{-}$) with some considerable ($\sim20\%$)
contribution of annihilation into quark-antiquark pair. In contrast,
for $\kappa_{\textrm{123}}=-0.4$ and $\kappa_{H2}=10^{-4}$, the
$N_{\textrm{DM}}$ annihilation processes into quark-antiquark pairs
continue being the most important. Moreover, $N_{\textrm{DM}}$ annihilation
processes into $JJ$ start to be considerable ($\sim15\%$). Similar
conclusions are true for the case of $\kappa_{123}=-0.1$ and $\kappa_{H2}=10^{-1}$.
However, for this case, annihilations into gauge bosons ($W^{\pm}/Z_{1}$)
have a little lower contribution when compared to the case $\kappa_{123}=-0.4$.
For the case of $\kappa_{123}=-0.1$ and $\kappa_{H2}=10^{-4}$, the
annihilations into antiquarks-quarks are the most contributing. 

In the region $120\,\textrm{GeV}\leq M_{\textrm{DM}}\leq180\,\textrm{GeV}$
and with $\kappa_{\textrm{123}}=-0.4$ and $\kappa_{H2}=10^{-1}$,
roughly speaking, three $N_{\textrm{DM}}$ annihilation processes
are similarly predominant. These are annihilations into $W^{+}W^{-}/Z_{1}Z_{1}$,
$R_{1}R_{1}$ and $JJ$. Recall that $R_{1}$ is the Higgs-like scalar.
For this region of mass and with $\kappa_{\textrm{123}}=-0.1$ and
$\kappa_{H2}=10^{-1}$ analogous conclusions can be reached. This
is not the case for $\kappa_{H2}=10^{-4}$ (with $\kappa_{\textrm{123}}=-0.4$)
in the same $M_{\textrm{DM}}$ region, in which case, $N_{\textrm{DM}}$
annihilations into $JJ$ are almost completely dominant with an additional
contribution ($\sim12\%$) from the annihilations into quark-antiquark
pairs. For $\kappa_{\textrm{123}}=-0.1$ and $\kappa_{H2}=10^{-4}$,
however, we find that annihilation into quark-antiquark pairs are
still dominant. 

As $M_{\textrm{DM}}$ is around $m_{R_{3}}/2\approx205$ GeV, we see
a resonance, in Fig. \ref{fig:Omega1} on the left, due $R_{3}$ $s-$channel
interchange for both $\kappa_{H2}$ values. Here, predominant annihilation
process is $N_{\textrm{DM}}N_{\textrm{DM}}\rightarrow JJ$ with more
than $50\%$ of contribution. It is also important to note that for
a $M_{\textrm{DM}}$ in this region we have the $\Omega_{\textrm{DM}}h^{2}$
value reported by Planck. This resonance does not occur in the $\kappa_{123}=-0.1$
cases because $m_{R_{3}}\approx1114.61$ GeV. However, in these cases,
we have one resonance at $m_{I_{2}}/2\approx242.6$ GeV, with $N_{\textrm{DM}}N_{\textrm{DM}}\rightarrow R_{1}J$
as dominant process for $\kappa_{H2}=10^{-1}$, and with $N_{\textrm{DM}}N_{\textrm{DM}}\rightarrow\bar{q}q$
as dominant process, for $\kappa_{H2}=10^{-4}$. 

In the region $220\,\textrm{GeV}\leq M_{\textrm{DM}}\leq500\,\textrm{GeV}$
and with $\kappa_{\textrm{123}}=-0.4$ and $\kappa_{H2}=10^{-1}$,
we can say that two annihilations, $N_{\textrm{DM}}N_{\textrm{DM}}\rightarrow JJ$
and $N_{\textrm{DM}}N_{\textrm{DM}}\rightarrow R_{3}J$, strongly
control $\Omega_{\textrm{DM}}h^{2}$. Except when $M_{\textrm{DM }}\approx m_{I_{2}}/2\approx485.1\,\textrm{GeV}$
where $N_{\textrm{DM}}N_{\textrm{DM}}\rightarrow R_{3}J$ annihilation
completely governs $\Omega_{\textrm{DM}}h^{2}$. As $500\,\textrm{GeV}<M_{\textrm{DM}}\leq700\,\textrm{GeV}$,
annihilations into $JJ$, $R_{1}I_{1}$, $R_{3}R_{3}$, $R_{3}J$
and $JI_{1}$ are predominant and their contributions depend on the
proximity to the three different resonances. Finally, when $700\,\textrm{GeV}<M_{\textrm{DM}}\leq1000\,\textrm{GeV}$,
annihilations into $JJ$, $R_{3}I_{1}$, $R_{3}J$, $N_{1}N_{1}$
and $N_{2}N_{2}$ are the most contributing processes to determine
$\Omega_{\textrm{DM}}h^{2}$. Similar behavior is found for the case
$\kappa_{\textrm{123}}=-0.4$ and $\kappa_{H2}=10^{-4}$. It is so
because in these regions of masses the annihilation processes mostly
depend on the trilinear vertices between scalars. 

As $\kappa_{\textrm{123}}=-0.1$ and $\kappa_{H2}=10^{-1}$, the scalar
spectrum changes and thus the location of the resonances change as
well. As it was commented, the resonance at $m_{R_{3}}/2\approx205\,\textrm{GeV}$
does not exist anymore. Instead, we have a resonance at $m_{I_{2}}/2\approx242.6$
GeV. In the region $220\,\textrm{GeV}\leq M_{\textrm{DM}}\leq470\,\textrm{GeV}$,
the most important difference, in contrast with the case $\kappa_{\textrm{123}}=-0.4$,
is that we do not have regions with $\Omega_{\textrm{DM}}h^{2}\leq0.119$
(a little tiny region can be seen in the $m_{I_{2}}/2\approx242.6\,\textrm{GeV}$).
Another difference is that annihilation into $W^{+}W^{-}/Z_{1}Z_{1}$
remains to be important in this region ($\sim15\%-35\%$). In addition,
annihilation into $R_{1}I_{1}$ contributes $>35\%$ in most of this
mass region. Other annihilations, such as $R_{1}J$, $I_{1}I_{1}$,
$JJ$ and $R_{1}R_{1}$, also contribute but are subdominant. For
$470\,\textrm{GeV}\leq M_{\textrm{DM}}\leq700\,\textrm{GeV}$, $\Omega_{\textrm{DM}}h^{2}$
is completely determined by annihilation into $I_{1}I_{1}$, $I_{1}J$,
$JJ$. As $700\,\textrm{GeV}\leq M_{\textrm{DM}}\leq1000\,\textrm{GeV}$,
annihilation processes into $N_{1}N_{1}$ and $N_{2}N_{2}$ share
importance with $I_{1}I_{1}$, $R_{3}I_{1}$, $R_{4}I_{1}$ and $R_{5}I_{1}$
to determine $\Omega_{\textrm{DM}}h^{2}$. For $M_{\textrm{DM}}>470\,\textrm{GeV}$,
we have $\Omega_{\textrm{DM}}h^{2}\leq0.119$.

Finally, when $\kappa_{\textrm{123}}=-0.1$ and $\kappa_{H2}=10^{-4}$,
we have some relevant differences. What is clearest is that for $M_{\textrm{DM}}<470\,\textrm{GeV}$
we just have the $Z_{1}$ resonance, which depends only on the VEVs
and the $g$'s. This is because of the smallness of $\kappa_{\textrm{123}}$
and $\kappa_{H2}$ (specially $\kappa_{H2}=10^{-4}$), which makes
the couplings with these $CP-$odd scalar mediators be tiny. Some
features are worth mentioning, though. Up to $M_{\textrm{DM}}\simeq350\,\textrm{GeV}$,
annihilation into quarks are predominant. After that, until $M_{\textrm{DM}}\simeq700\,\textrm{GeV}$,
the final products $I_{1}I_{1}$, $I_{1}J$, $JJ$ (summing $\sim35-100\%$)
enter as the major contributors to the relic density and the quarks
enter as subdominant processes fading out at $M_{\textrm{DM}}\simeq450\,\textrm{GeV}$.
Next, up to $M_{\textrm{DM}}\simeq900\,\textrm{GeV}$, the main annihilation
products are $N_{1}N_{1}$ and $N_{2}N_{2}$ ($\sim30-40\%$ each),
with $R_{3}I_{1}$ taking place at the end of this interval. Finally,
for $850\,\textrm{GeV}\leq M_{\textrm{DM}}\leq1000\,\textrm{GeV}$,
the main contributions come from $R_{3}I_{1}$, $R_{4}I_{1}$ and
$R_{5}I_{1}$, summing more than $70\%$ of the DM annihilation energy. 

\begin{figure}
\includegraphics[scale=1.1]{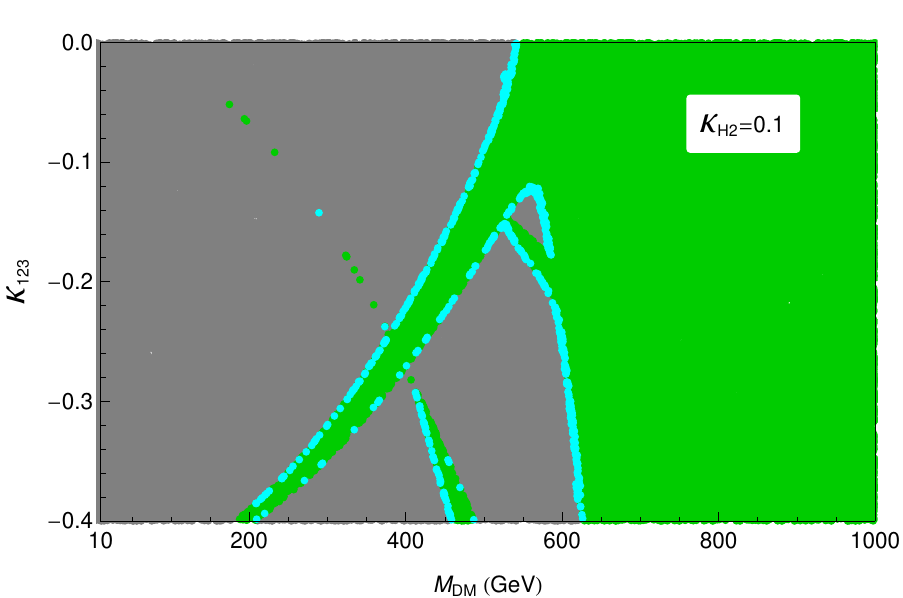}

\protect\caption{(Color online) 2-D figure displaying the behavior of $\Omega_{\textrm{DM}}h^{2}$
as one continuously varies both $\kappa_{\textrm{123}}$ in the range
$[-0.4,\,0]$, with $\kappa_{H\textrm{2}}=0.1$, and $M_{\textrm{DM}}$
in the range $[10,\,1000]\,$ GeV. The cyan points represent correct
relic density $\Omega_{\textrm{DM}}h^{2}$, within experimental errors;
the green ones indicate $\Omega_{\textrm{DM}}h^{2}$ below the Planck
result; and the gray ones mean $\Omega_{\textrm{DM}}h^{2}$ above
Planck constraint and thus ruled out.\label{fig:2D_Omega}}
\end{figure}

Now, in order to grasp the behavior of the relic density when one
varies $\kappa_{\textrm{123}}$, we show a two-dimensional figure,
Fig. \ref{fig:2D_Omega}, which was obtained with ${\tt MicrOMEGAs}$,
from a $10^{5}$ points iteration. We see from it that as one varies
$\kappa_{\textrm{123}}$, the regions for correct relic density (cyan
points), change place, getting to the minimal value of $M_{\textrm{DM}}\sim200\textrm{ GeV}$
for $\kappa_{\textrm{123}}=-0.4$; and also for some points which
increases in $\kappa_{123}$ as $M_{\textrm{DM}}$ decreases, having
at $\kappa_{\textrm{123}}\sim-0.05$ its last point. We can also notice
green regions (together with cyan lines) that extend from left to
right as $M_{\textrm{DM}}$ increases, and the reason behind it are
the resonances of $I_{2}$, $I_{3}$ (decrease as $\kappa_{\textrm{123}}$
increases) and $R_{3}$ (increases). Therefore, one can conclude that
the correct relic density, before $M_{\textrm{DM}}\sim500-600\,\textrm{GeV}$,
may only be reached through resonances of the lightest singlet particles
of our spectrum.

\subsection{Direct Detection}

\begin{figure}
\includegraphics[scale=0.4]{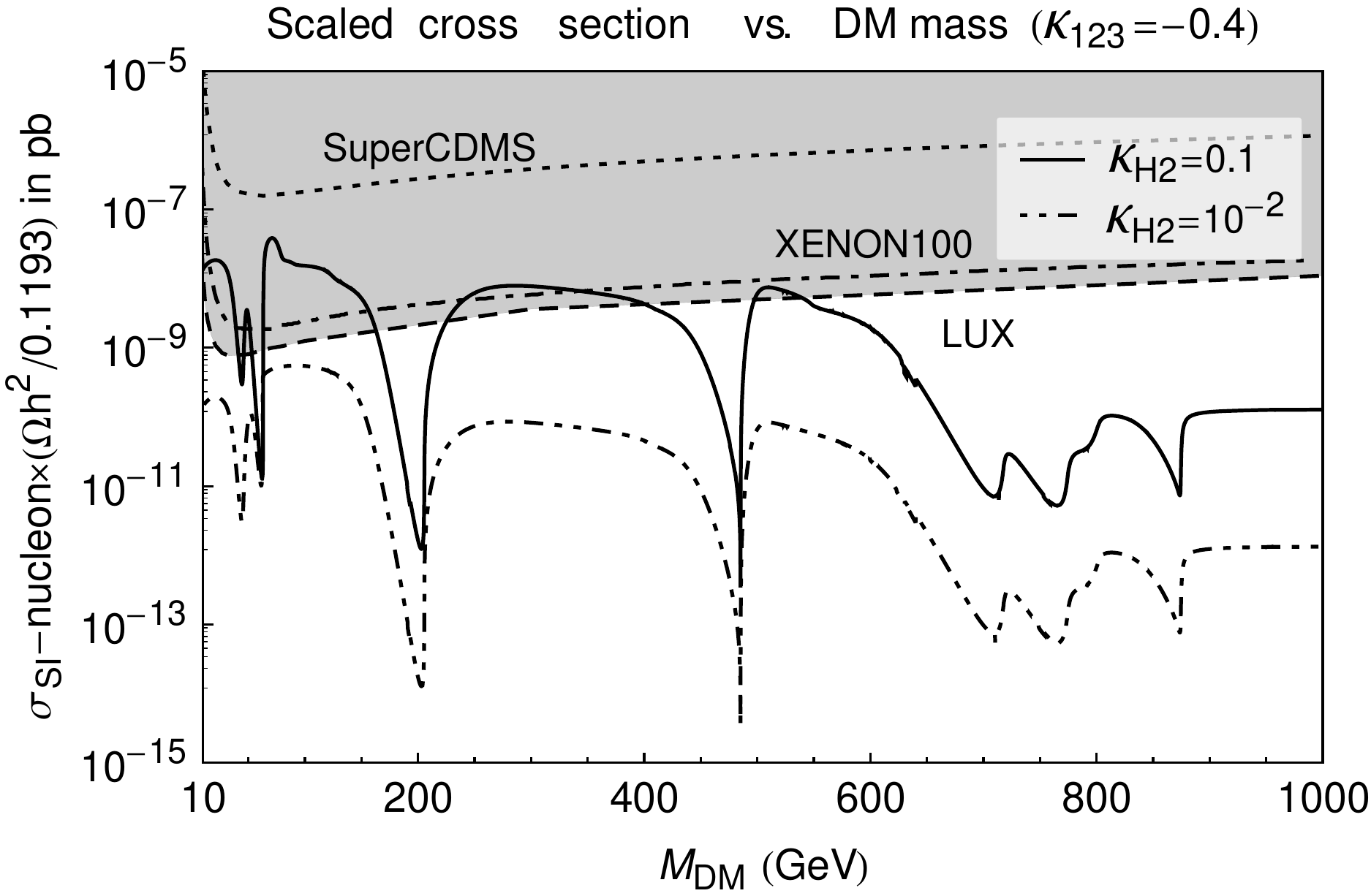}\includegraphics[scale=0.4]{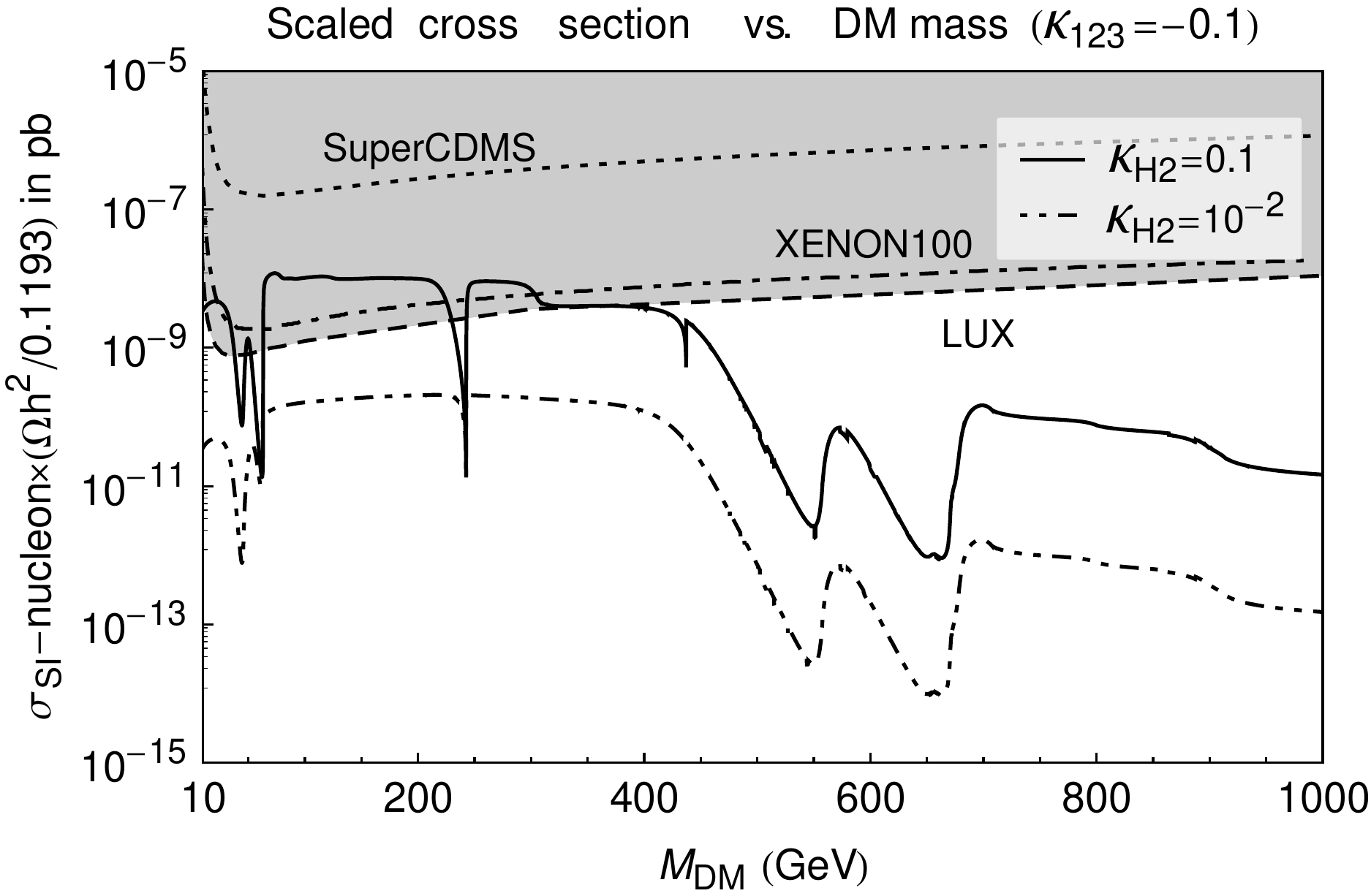}

\protect\caption{Figures displaying the curves representing the SI cross section per
nucleon, $\sigma_{\textrm{SI-nucleon}}$, asa function of $M_{\textrm{DM}}$,
for the $N_{\textrm{DM}}$ elastic scattering off nucleon. Cases for
$\kappa_{\textrm{123}}=-0.4;\,-0.1$ and $\kappa_{H\textrm{2}}=10^{-1},10^{-2}$
are shown. In these figures, we also display the SI upper limits coming
from LUX (dashed), XENON100 (dot-dashed) and SuperCDMS (dotted). All
$\sigma_{\textrm{SI-nucleon}}$ curves in the gray region are ruled
out by the LUX upper limit. \label{fig:DDetection}}
\end{figure}

Other important constraints on DM candidates come from the current
experiments \citep{XENON100_2012,SuperCDMS_2014,LUX_2014} which aim
to directly detect WIMP dark matter by measuring the kinetic energy
transferred to a nucleus after it scatters off a DM particle. All
of these experiments have imposed limits on the WIMP scattering cross
section off the nuclei. In general, the WIMP-nucleus interactions
can be either spin-independent (SI) or spin-dependent (SD). Currently,
the most constraining limits come from the Large Underground Xenon
(LUX) experiment \citep{LUX_2014} which has set bounds on the SI
WIMP-nucleon elastic scattering with a minimum upper limit on the
cross section of $7.6\times10^{-10}$ pb at a WIMP mass of $33$ GeV$/c^{2}$. 

We have verified that, for $N_{\textrm{DM}}$ considered here, the
dominant interactions are SI. Thus, we calculate (using the ${\tt MicrOMEGAs}$
package) the SI elastic scattering cross section per nucleon, $\sigma_{\textrm{SI-nucleon}}$,
and the results are shown in Fig. \ref{fig:DDetection}. Actually,
we scale the $\sigma_{\textrm{SI-nucleon}}$ cross sections with the
calculated relic density relative to that measured by the Planck in
order to properly compare the predicted cross sections with those
given by direct detection experiments, which present their results
assuming the observed density. The experimental limits on SI cross
sections are also shown in Fig. \ref{fig:DDetection}. We have not
shown results for SD cross sections because we found that those are,
in general, several orders under the SI current limits, see Refs.
\citep{COUPP_SD_DD,SIMPLE-II_SD_DD} which state a minimum upper bound
of $\sim5\times10^{-3}\,\textrm{pb}$ at a WIMP mass of $24\,\textrm{GeV}/c^{2}$. 

From Fig. \ref{fig:DDetection}, it can be seen that the smaller $\kappa_{H2}$,
the smaller $\sigma_{\textrm{SI-nucleon}}$. As $\kappa_{H2}=10^{-2}$
the $\sigma_{\textrm{SI-nucleon}}$ is below the LUX upper bound for
all values of $M_{\textrm{DM}}$. For $\kappa_{H2}=10^{-1}$ and $M_{\textrm{DM}}\lesssim500$
GeV, $\sigma_{\textrm{SI-nucleon}}$ is below the LUX limit only around
the resonances. In contrast, for $M_{\textrm{DM}}\gtrsim500$ GeV,
the LUX limits are satisfied for all cases shown in Fig. \ref{fig:DDetection}.
This implies that $\sigma_{\textrm{SI-nucleon}}$ mainly depends on
$\kappa_{H2}$. This fact is easily understood by realizing that,
in our case, the relevant interactions for direct detection are mostly
mediated via Higgs in the t-channel . Thus, these interactions depend
on the mixings between the Higgs scalar ($R_{1}$) and rest of $R_{i}$
scalars. These mixings strongly depend on the $\kappa_{H2}$ value,
as it was already discussed. In addition, we can see from Fig. \ref{fig:DDetection}
that although $\sigma_{\textrm{SI-nucleon}}$ (actually $\sigma_{\textrm{SI-nucleon}}\times\Omega_{\textrm{DM}}h^{2}/0.1193$
) does not depend directly on other scalars, there is clearly indirect
dependence on them because these scalars affect the annihilation cross
section and thus the relic abundance. 

Finally, from Figs. \ref{fig:Omega1} and \ref{fig:DDetection} we
can conclude that, provided $\kappa_{H2}\lesssim10^{-2}$, the constraints
coming from the $\Omega_{\textrm{DM}}h^{2}$ determine if a set of
parameters leads to a viable dark matter candidate or not.

\section{conclusions}

\label{sec:Conclusions}In this paper, we have discussed a scenario
where neutrino masses and dark matter are possible. In particular,
the model presented here is a gauge extension of the SM based on $\text{SU}(2)_{L}\otimes\text{U}(1)_{Y^{\prime}}\otimes\text{U}(1)_{\mathcal{B-L}}$
symmetry group. Besides the SM fields, we have added one doublet scalar
($\Phi$), four singlet scalars ($\phi_{i}$), and three right-handed
neutrinos ($n_{Ri}$). These three last fields have different quantum
numbers under the gauge groups and more importantly, they couple to
different scalars. This allows a rich texture in the neutrino mass
matrices. In addition, because of the exclusion of one term from the
Lagrangian, we have that a $\ZZ_{2}$ symmetry acting only in $n_{R3}$
appears. This opens the possibility that $n_{R3}$ be a DM candidate. 

The model contains a very rich scalar sector, and in special, we find
that it contains a Majoron, $J$, which has its origin in a global
accidental symmetry, U$\left(1\right)_{J}$. We show that this symmetry
is exact and it acts on the total Lagrangian. Despite the fact that
global symmetries can be broken by gravity quantum effects, we have
not considered that possibility in this paper. Thus, the Majoron remains
massless and in principle, it poses some issues to the safety of the
model. Therefore, we study the consequences of the presence of $J$
in the physical spectrum. Specifically, we consider four major challenges:
(i) energy loss in stars by the process $\gamma+e^{-}\rightarrow e^{-}+J$
\citep{geJ_stars1,geJ_stars2}; (ii) relativistic degrees of freedom
in the Universe, parametrized by $N_{\textrm{eff}}$; (iii) $Z$ invisible
decay width; and (iv) Higgs invisible decay width. Setting $V_{\phi}=1$
TeV and $V_{H}\simeq V_{\textrm{SM}}=246$ GeV, the first of these
constraints leads to $\epsilon=V_{\Phi}/V_{\phi}\lesssim3.8\times10^{-4}$
(where we have set all VEVs of scalar singlets equal). The second
constraint really does not lead to any restriction on the parameters
of the model since the $J$ contribution to the density of radiation
in the Universe is in agreement with the Planck limits \citep{PLANCK_2013}
for different decoupling temperatures. The issue of the $Z$ invisible
width is also overcome by imposing some constraints on the scalar
potential parameters as shown in Secs. \ref{sec:Scalar Sector} and
\ref{sec:dark_matter}. In addition, the Higgs invisible decay width
is maintained at $\textrm{Br}_{\textrm{Higgs}}^{\textrm{Inv}}\apprle10-15\%$
\citep{Ellis_2013,Giardino2014,Atlascollaboration,CMScollaboration}
provided that $\kappa_{H2}<0.2$. Finally, since $\epsilon\ll1$,
we have shown an overview of the scalar spectra by expanding the scalar
squared-mass matrices in powers of $\epsilon$ and doing other simplifying
considerations.

Since $\epsilon$ has to be very small, we can use the well-known
see-saw approximation to analytically solve the ${\cal D}_{ij}$ and
$\mathcal{M}_{ij}$ parameters. These are found by imposing some experimental
constraints coming from the neutrino physics. In special, the mixing
angles and the differences of the squared neutrino masses. We manage
to solve the ${\cal D}_{ij}$ and $\mathcal{M}_{ij}$ parameters by
making the ansatz that $M_{\nu}$ matrix is diagonalized by the tri-bimaximal-Cabbibo
matrix. It is important to note that the existence of the $\ZZ_2$
symmetry makes easier to solve the equations because there appears
a massless light neutrino. In general, we find all ${\cal D}_{ij}$
parameters depend on the dimensional constant $K=\frac{V_{\phi}^{2}\epsilon^{2}}{2M_{N}}$.
It is true for both normal and inverted mass hierarchies. One more
interesting result is reached when we take into consideration the
LFV processes such as $\textrm{Br}\left(\mu\rightarrow e+\gamma\right)$
and $\textrm{Br}\left(\tau\rightarrow\mu+\gamma\right)$. The current
bounds \citep{Olive:2014PDG} on these processes constrain $\epsilon$.
For the normal case, one obtains $\epsilon\apprge1.97\times10^{-6}$,
and for the inverted case $\epsilon\gtrsim1.58\times10^{-6}$. We
also have checked that $\sum_{i=1}^{3}m_{\nu_{i}}<0.23\,\textrm{eV}$,
coming from Planck \citep{PLANCK_2013}, and the effective Majorana
mass bound $m_{ee}<0.27-0.65\,\textrm{GeV}$ \citep{mee_future,mee_CUORE_2015},
coming from double beta decay experiments, were satisfied.

After the scalar and neutrino sectors of the model were studied and
many of the parameters were set, we consider the $n_{R3}$ (more precisely
$N_{\textrm{DM}}$) as a DM candidate. We study the bounds coming
from the relic density abundance $\Omega_{\textrm{DM}}h^{2}$ \citep{PLANCK_2013}
and the direct detection experiments \citep{LUX_2014,XENON100_2012,SuperCDMS_2014}.
Basically, we have worked with three free parameters, $\kappa_{H2}$
, $\kappa_{123}$, and the DM mass, $M_{\textrm{DM}}$. These parameters
have been chosen because they play a very important role in determining
both the $N_{\textrm{DM}}$ annihilation cross section and the $N_{\textrm{DM}}$
elastic scattering off the nucleon. Roughly speaking, we find that
for $M_{\textrm{DM}}\lesssim500$ GeV, the $\Omega_{\textrm{DM}}h^{2}\leq0.1193$
is achieved around the resonance regions. For $M_{\textrm{DM}}>500$
GeV there are several regions with $\Omega_{\textrm{DM}}h^{2}\leq0.1193$
aside from resonances regions. This is understood by realizing that
the couplings of $N_{\textrm{DM}}$ to scalars (including the Higgs)
are proportional to $\frac{\sqrt{2}M_{\textrm{DM}}}{2V_{\phi}}$,
and that for $M_{\textrm{DM}}<200$ GeV the main annihilation channels
are, in general, mediated by the Higgs. It is also observed, that
making $\kappa_{123}$ bigger, we obtain more regions with $\Omega_{\textrm{DM}}h^{2}\leq0.1193$
for smaller $M_{\text{DM}}$. This is because $\kappa_{123}$ strongly
controls the trilinear couplings between scalars and thus, the annihilation
cross section is larger when $\kappa_{123}$ is larger. We have found
that the relative contributions to the DM annihilation in this model
have an intricate pattern. It strongly depends on the scalar masses.
However, some general conclusions can be drawn. For $M_{\textrm{DM }}\leq200$
GeV the annihilation into $\bar{q}q$, $W/Z$, $\bar{l}l$ are dominant.
For $200<M_{\textrm{DM}}<700$ GeV annihilations into scalars are
the most important. Finally, for $700\,\text{GeV}<M_{\textrm{DM}}$
annihilations into $N_{i}N_{i}$ play a important role. 

For DM direct detection, the parameter $\kappa_{H2}$ is the most
relevant since it is the only one which effectively couples $N_{\textrm{DM}}$
to the quarks in our model. Since nuclei are made of quarks (and gluons),
this interaction is of supreme importance to the elastic scattering
of $N_{\textrm{DM}}$ off nuclei. We found out that if we choose $\kappa_{H2}=10^{-2}$,
our entire curves are below LUX data, the most stringent upper bounds
on SI DD, however if one chooses a riskier value such as $\kappa_{H2}=0.1$,
it can still be lower than LUX, however only above $M_{\textrm{DM}}\sim500-600\,\textrm{GeV}$
or at the resonances below that $M_{\textrm{DM}}$.

A final remark concerning to the $Z_{2}$ gauge boson is in order.
In the region of parameters that we have studied, the $Z_{2}$ boson
does not affect the DM properties. It is because $Z_{2}$ is heavy
since its mass has to satisfy $M_{Z_{2}}/g_{B-L}\apprge6\,\textrm{TeV}$
\citep{MZ2_1,MZ2_2}. In addition, its mixing angle in the neutral
current is limited to be $\tan\beta\apprle10^{-3}$ \citep{beta_1,beta_2,beta_3}. 
\begin{acknowledgments}
B. L. S. V. would like to thank Coordenação de Aperfeiçoamento de
Pessoal de Nível Superior (CAPES), Brazil, for financial support under
contract 2264-13-7 and the Argonne National Laboratory for kind hospitality.
E. R. S. would like to thank Conselho Nacional de Desenvolvimento
Científico e Tecnológico (CNPq), Brazil, for financial support under
process 201016/2014-1, and Bethe Center for Theoretical Physics and
Physikalisches Institut, Universität Bonn, for warm hospitality.
\end{acknowledgments}

\appendix

\section{THE MINIMIZATION \label{Appendix}}

The general minimization conditions coming from $\partial V_{{\cal {B}-{\cal {L}}}}/\partial R_{i}=0$,
where $V_{{\cal {B}-{\cal {L}}}}$ is the scalar potential in Eq.
(\ref{potential}) and $R_{i}=\{H_{R}^{0},\,\Phi_{R}^{0},\,\phi_{1R},\,\phi_{2R},\,\phi_{3R},\,\phi_{XR}\}$
are the neutral real components of the scalar fields, can be written
as:
\begin{eqnarray}
0 & = & V_{H}\left(2\lambda_{H}V_{H}^{2}+\kappa_{H\Phi}V_{\Phi}^{2}+\kappa_{H1}V_{\phi_{1}}^{2}+\kappa_{H2}V_{\phi_{2}}^{2}+\kappa_{H3}V_{\phi_{3}}^{2}+\kappa_{HX}V_{\phi_{X}}^{2}-2\mu_{H}^{2}\right)\nonumber \\
 &  & -\sqrt{2}\kappa_{H\Phi X}V_{\Phi}V_{\phi_{X}};\label{v1}\\
0 & = & V_{\Phi}\left(\kappa_{H\Phi}V_{H}^{2}+2\lambda_{\Phi}V_{\Phi}^{2}+\kappa_{\Phi1}V_{\phi_{1}}^{2}+\kappa_{\Phi2}V_{\phi_{2}}^{2}+\kappa_{\Phi3}V_{\phi_{3}}^{2}+\kappa_{\Phi X}V_{\phi_{X}}^{2}-2\mu_{\Phi}^{2}\right)\nonumber \\
 &  & -\sqrt{2}\kappa_{H\Phi X}V_{H}V_{\phi_{X}};\label{v2}\\
0 & = & V_{\phi_{1}}\left(\kappa_{H1}V_{H}^{2}+\kappa_{\Phi1}V_{\Phi}^{2}+2\lambda_{1}V_{\phi_{1}}^{2}+\kappa_{12}V_{\phi_{2}}^{2}+\kappa_{13}V_{\phi_{3}}^{2}+\kappa_{1X}V_{\phi_{X}}^{2}-2\mu_{1}^{2}\right)\nonumber \\
 &  & +V_{\phi_{2}}V_{\phi_{3}}\left(\kappa_{123}V_{\phi_{3}}+\kappa_{123X}V_{\phi_{X}}\right);\\
0 & = & V_{\phi_{2}}\left(\kappa_{H2}V_{H}^{2}+\kappa_{\Phi2}V_{\Phi}^{2}+\kappa_{12}V_{\phi_{1}}^{2}+2\lambda_{2}V_{\phi_{2}}^{2}+\kappa_{23}V_{\phi_{3}}^{2}+\kappa_{2X}V_{\phi_{X}}^{2}-2\mu_{2}^{2}\right)\nonumber \\
 &  & +V_{\phi_{1}}V_{\phi_{3}}\left(\kappa_{123}V_{\phi_{3}}+\kappa_{123X}V_{\phi_{X}}\right);\\
0 & = & V_{\phi_{3}}\left(\kappa_{H3}V_{H}^{2}+\kappa_{\Phi3}V_{\Phi}^{2}+\kappa_{13}V_{\phi_{1}}^{2}+\kappa_{23}V_{\phi_{2}}^{2}+2\lambda_{3}V_{\phi_{3}}^{2}+\kappa_{3X}V_{\phi_{X}}^{2}+3\kappa'_{3X}V_{\phi_{3}}V_{\phi_{X}}\right.\nonumber \\
 &  & \left.-2\mu_{3}^{2}\right)+V_{\phi_{1}}V_{\phi_{2}}\left(2\kappa_{123}V_{\phi_{3}}+\kappa_{123X}V_{\phi_{X}}\right);\\
0 & = & V_{\phi_{X}}\left(\text{\ensuremath{\kappa_{HX}}}V_{H}^{2}+\text{\ensuremath{\kappa_{\Phi X}}}V_{\Phi}^{2}+\ensuremath{\kappa_{1X}}V_{\phi_{1}}^{2}+\text{\ensuremath{\kappa_{2X}}}V_{\phi_{2}}^{2}+\text{\ensuremath{\kappa_{3X}}}V_{\phi_{3}}^{2}+2\text{\ensuremath{\lambda_{X}}}V_{\phi_{X}}^{2}-2\mu_{X}^{2}\right)\nonumber \\
 &  & -\sqrt{2}\text{\ensuremath{\kappa_{H\Phi X}}}V_{\Phi}V_{H}+V_{\phi_{3}}\left(\kappa_{123X}V_{\phi_{1}}V_{\phi_{2}}+\kappa'_{3X}V_{\phi_{3}}^{2}\right).\label{v6}
\end{eqnarray}

\bibliographystyle{unsrt}
\bibliography{references}

\end{document}